\documentclass[lettersize,journal]{IEEEtran}

\usepackage{amsmath,amsfonts}
\usepackage{algorithmic}
\usepackage{algorithm}
\usepackage{array}
\usepackage{makecell}
\usepackage{subcaption}
\usepackage[caption=false,font=normalsize,labelfont=sf,textfont=sf]{subfig}
\usepackage{textcomp}
\usepackage{stfloats}
\usepackage{url}
\usepackage{verbatim}
\usepackage{graphicx}
\usepackage{cite}
\usepackage{booktabs}
\usepackage{tabularray}
\usepackage{xcolor}
\usepackage{tabularx}
\usepackage{multirow}
\usepackage{siunitx}
\usepackage{threeparttable}

\begin{document}

\title{A Comprehensive Study on the Effectiveness of ASR Representations for Noise-Robust Speech Emotion Recognition}

\author{Xiaohan Shi, Jiajun He, Xingfeng Li, Tomoki Toda,~\IEEEmembership{Senior Member,~IEEE}
        % <-this % stops a space
\thanks{This work was financially supported by JST SPRING, Grant Number JPMJSP2125, and in part by JST AIP Acceleration Research JPMJCR25U5, Japan, and JSPS KAKENHI Grant Number 21H05054.}% <-this % stops a space
\thanks{Xiaohan Shi is with the Graduate School of Informatics, Nagoya University, Nagoya 464-8601, Japan (e-mail: xiaohan.shi@g.sp.m.is.nagoya-u.ac.jp)}
\thanks{Jiajun He is with the Graduate School of Informatics, Nagoya University, Nagoya 464-8601, Japan (e-mail: Jiajun.he@g.sp.m.is.nagoya-u.ac.jp)}
\thanks{Xingfeng Li is with the Faculty of Data Science, City University of Macau, Macau, China (e-mail: xfli@cityu.edu.mo)}
\thanks{Tomoki Toda is with the Information Technology Center, Nagoya University, Nagoya 464-8601, Japan (e-mail: tomoki@icts.nagoya-u.ac.jp)}
\thanks{Manuscript received April 19, 2021; revised August 16, 2021.}}

\markboth{Journal of \LaTeX\ Class Files,~Vol.~14, No.~8, August~2021}%
{Shell \MakeLowercase{\textit{et al.}}: A Sample Article Using IEEEtran.cls for IEEE Journals}


\maketitle

\begin{abstract}
In this paper, we propose an efficient noise-robust approach to noisy speech emotion recognition (NSER). Although conventional NSER methods effectively handle some types of noise, such as stationary noise, they struggle with more complex noise encountered in realistic acoustic environments owing to its complexity and unpredictability. To address this issue, we introduce a novel NSER method that leverages automatic speech recognition (ASR) models as noise-robust feature extractors, filtering out non-vocal information from noisy speech. Specifically, we extract intermediate layer representations from the ASR model to capture emotional speech features for the NSER task. In this study, we conduct a comprehensive evaluation of ASR representations against traditional high-level statistical function (HSF) features, noise reduction approaches, mainstream NSER models, and fine-tuned self-supervised learning (SSL) methods. We analyze encoder and decoder groups through both single-layer and fusion-based multi-layer representations, further investigating the layer-wise performance and robustness of encoder and decoder modules under different noise types and intensities. Additionally, we explore their correlation with ASR performance, assess cross-modal performance using ASR transcripts, and examine cross-lingual generalization. Our experimental results demonstrate that (1) the proposed method outperforms HSF features, noise reduction approaches, mainstream NSER models, and SSL methods; (2) the adapter-based method further enhances ASR representations; (3) higher encoder and lower decoder layers (e.g., in Whisper) yield the best individual performance, while combining encoder and decoder layers across all depths achieves the strongest results; (4) performance decreases as noise intensity increases, especially with human speech noise; (5) this decline mirrors the degradation observed in ASR performance as noise intensity increases; (6) our approach surpasses transcript-based methods using ASR or ground-truth transcriptions; and (7) it achieves robust cross-lingual performance compared with mainstream SSL representations.
\end{abstract}


\section{Introduction}
\IEEEPARstart{S}{peech} emotion recognition (SER) has attracted considerable attention over the past two decades, particularly within the domain of human-computer interaction \cite{schuller2018speech, cowie2001emotion, brave2007emotion}. SER has enabled a variety of applications, fundamentally transforming customer service interactions by allowing chatbots to provide empathetic and contextually appropriate responses \cite{deschamps2021end, ghandeharioun2019emma}. Additionally, SER has been integrated into in-vehicle dashboard systems, improving driver experience and safety through real-time understanding and responsiveness to emotional states \cite{harris2011emotion, zepf2020driver}. Moreover, the utility of this technology extends to speech-to-speech translation platforms, enhancing the accuracy and naturalness of cross-linguistic communication by incorporating emotional context \cite{akagi2014toward, elbarougy2014toward}. Despite these promising applications, SER systems face a significant challenge: the prevalence of unknown noise in realistic acoustic environments. This noise severely hampers the performance of SER models, preventing them from effectively distinguishing and accurately recognizing emotions amidst environmental disturbances. Such limitations pose a considerable barrier to the widespread adoption and effective implementation of SER across diverse domains \cite{you2006emotion, schuller2006emotion}.

Empirical investigations have demonstrated the efficacy of established methodologies in mitigating common noise sources, such as white Gaussian noise, in speech-related tasks \cite{tiwari2020multi, pandharipande2018unsupervised, murata2001approach, chenchah2017bio, sekkate2019investigation, nam2021cascaded}. However, translating such successes to realistic acoustic environments remains challenging due to a distinct category of noise, including ambient sounds such as the clicking of high-heeled shoes or abrupt door knocking. Unlike Gaussian noise, these environmental noises exhibit non-Gaussian distributions and inherent unpredictability, complicating their effective reduction or elimination through conventional noise reduction techniques \cite{you2006emotion, heracleous2017speech}. Consequently, the performance of SER systems, already vulnerable to noise, faces greater limitations in accurately recognizing emotional cues under such disturbances \cite{basu2017review}. Therefore, effectively adapting SER systems to complex acoustic environments is crucial for enhancing their robustness and applicability across diverse contexts.

Automatic speech recognition (ASR) serves as a cornerstone in contemporary communication systems, facilitating the conversion of spoken language into written text with exceptional precision \cite{benzeghiba2007automatic}. Its applicability spans a broad spectrum of domains, including virtual assistants, transcription services, and voice-controlled devices, among others \cite{jefferson2019usability}. Central to ASR models is their proficiency in discerning and interpreting human speech amidst diverse environmental challenges, such as background noise and reverberations \cite{li2015robust}. Notably, ASR systems are meticulously engineered to differentiate between vocal and non-vocal elements of speech, thereby enhancing their robustness against noise interference \cite{vipperla2010ageing}. Thus, we expect that ASR can effectively serve as a noise-robust feature extractor by prioritizing the extraction of vocal features while attenuating extraneous acoustic signals.

Capitalizing on this intrinsic capability, we propose an innovative approach to address the challenge of noisy speech emotion recognition (NSER) in realistic acoustic environments. In our framework, ASR representations play a pivotal role in capturing fundamental vocal cues that underpin emotional expressions. By leveraging the inherent robustness of ASR-derived features to noise, we aim to mitigate the harmful effects of environmental disturbances on emotion recognition accuracy. First, we compare ASR representations with several strong baselines, including traditional high-level statistical function (HSF) features \cite{Schuller:2009, Schuller:2010, Schuller:2011, Schuller:2012, Schuller:2013, Eyben:2015}, conventional noise reduction methods \cite{luo2019conv, hu2020dccrn}, mainstream NSER models \cite{liu2023multi, liu2025enhanced}, and fine-tuned self-supervised learning (SSL) models \cite{baevski2020wav2vec, hsu2021hubert, chen2022wavlm}. Second, we examine the effectiveness of single-layer and fusion-based multi-layer ASR representations, where the single-layer setting adopts the last layer or the best-performing layer among all layers, and the fusion setting includes simple fusion (mean pooling), parameterized fusion (weighted sum and scalar mixing), and advanced fusion (attention-based pooling). Third, we conduct a layer-wise analysis to compare the performance of encoder and decoder representations, further contrasting them with SSL features. Fourth, we assess robustness to different noise intensities by testing under various signal-to-noise ratio (SNR) conditions. Fifth, we investigate the correlation between ASR and NSER performance by analyzing their degradation trends across noise conditions. Sixth, we evaluate the effectiveness of ASR representations in NSER by assessing their performance across different modalities. Finally, we assess the generalization of ASR representations by testing their effectiveness in cross-lingual NSER scenarios.

The contributions of this paper are summarized as follows:
\begin{itemize}
\item Introduction of a novel ASR-based approach for noise-robust NSER in realistic acoustic environments.
\item Proposal of a method for leveraging ASR encoder and decoder representations for NSER.
\item Conducting a systematic layer-wise analysis of ASR models, which reveals that higher encoder layers and lower decoder layers (e.g., Whisper) contribute most effectively to NSER.
\item Evaluation of the performance of ASR representations across different types and intensities of noise.
\item Exploration of the correlation between ASR performance and NSER outcomes.
\item Examination of the impact of ASR representations across various modalities, including a comparison between ASR and text modalities using SSL methods for transcriptions.
\item Investigation of the robustness of ASR representations in cross-lingual scenarios, assessing the robustness of the proposed models in diverse linguistic contexts.
\end{itemize}

The remaining sections of this paper are organized as follows: In Section II, we provide an overview of NSER and the application of ASR in SER, forming the foundation for the proposed approach. In Section III, we present the proposed method and model structure. In Section IV, we describe the three databases used in this study. In Section V, we outline the experimental setup, including the control group design, implementation details, and evaluation metrics. In Section VI, we present the experimental results, validating the effectiveness of the proposed model. Finally, in Section VII, we discuss findings and outline future work.

\section{Related Work}
\subsection{Noisy Speech Emotion Recognition}
In realistic acoustic environments, various types of noise arise, each with distinct characteristics and sources \cite{kim1999auditory}. Gaussian white noise is a typical example, representing a fundamental manifestation of signal randomness. It is characterized as stationary noise with a uniform distribution of power spectral density across all frequencies, symbolizing an essential element of signal variability \cite{marmarelis2012analysis}. Conversely, impulse noise consists of sudden, high-amplitude bursts of pulses or spikes, often arising from equipment malfunctions or electromagnetic interference incidents \cite{henderson1986impulse}. Additionally, environmental noise encompasses a broad spectrum of background disturbances from the surrounding environment, including sources such as vehicular noise, industrial machinery operations, and human activities, all contributing to environmental noise \cite{munzel2018environmental}.

To address the multifaceted challenges posed by diverse noise patterns in realistic acoustic environments, numerous studies have been conducted to mitigate the detrimental impact of noise on the accuracy of speech-related tasks, particularly in NSER. These endeavors typically employ three primary strategies: signal-level, feature-level, and model-level interventions, as elucidated by Tiwari et al. \cite{tiwari2020multi}. For example, Pandharipande et al. \cite{pandharipande2018unsupervised} proposed an unsupervised method that uses a front-end voice activity detector to selectively extract frames with higher SNRs from spoken utterances, aiming to enhance accuracy under noisy conditions by improving signal quality at the signal level. Similarly, Vasquez-Correa et al. \cite{vasquez2015emotion} proposed a method for emotion recognition using wavelet packet transform features. They classified negative emotions and distinguished them from neutral and positive states by analyzing voiced and unvoiced segments. Features such as log energy and mel-frequency cepstral coefficients were calculated, and the method was tested on the Berlin Emotional Database and the eNTERFACE05 Database under various noise conditions. They also evaluated two speech enhancement techniques, noting improved accuracy in noisy environments. At the feature level, Chenchah and Lachiri \cite{chenchah2017bio} introduced a novel approach involving MFCC-shifted-delta-cepstral coefficients. On the other hand, Sekkate et al. \cite{sekkate2019investigation} proposed an approach combining baseline mel-frequency cepstral coefficients, discrete wavelet transform-derived MFCCs, and pitch-based features to create a feature set that effectively captures relevant information in a relatively low-dimensional space. At the model level, Tiwari et al. \cite{tiwari2020multi} devised multi-conditioning and data augmentation using an utterance-level parametric generative noise model. This strategy aims to encompass the entire noise space in the mel-filterbank energy domain, rendering the model robust against unseen noise conditions. Additionally, Nam and Lee \cite{nam2021cascaded} introduced a novel approach employing a cascaded denoising CNN architecture for SER under noisy conditions. This architecture consists of two stages: DnCNN for denoising using residual learning and CNN for subsequent classification at the model level. More recently, Liu et al. \cite{liu2023multi} introduced a multi-level knowledge distillation framework that transfers robust knowledge across network depths, thereby enhancing SER resilience under noisy conditions. Building on this, Liu et al. \cite{liu2025enhanced} proposed an adaptive emotion denoising diffusion model combined with an iterative confidence learning strategy, achieving improved performance in noisy environments.

\subsection{Automatic Speech Recognition for Speech Emotion Recognition}
ASR is a critical tool for transcribing spoken language into written text, with broad applicability across various diverse application domains \cite{ten2003emotions}. Significant progress has been made in the field of SER through the use of ASR-generated transcripts. Yoon et al. \cite{yoon2018multimodal} pioneered a deep dual recurrent encoder model that simultaneously leveraged both audio signals and text data from the Google Cloud Speech API. Subsequently, Sahu et al. \cite{sahu2019multi} used transcripts from two commercial ASR systems for a bimodal SER approach (audio and text), noting a relative decline in unweighted accuracy compared with ground-truth transcripts. Li et al. \cite{li2020learning} introduced a temporal alignment mean-max pooling mechanism to capture subtle, fine-grained emotions within utterances, as well as a cross-modality excitement module to make sample-specific adjustments to embeddings. Santoso et al. \cite{santoso2021speech} proposed a confidence measure to adjust the importance weights of ASR transcripts based on the likelihood of recognition errors in each word, effectively mitigating the impact of ASR errors on SER performance. Wu et al. \cite{wu2021emotion} developed a dual-branch model to enhance robustness against ASR errors, featuring a time-synchronous branch that combines speech and text modalities, alongside a time-asynchronous branch that integrates sentence text embeddings from contextual utterances. Shon et al. \cite{shon2021leveraging} generated pseudo labels for ASR transcripts in semi-supervised speech sentiment analysis. Feng and Narayanan \cite{feng2024foundation} fused audio with ASR transcripts from an advanced ASR model, demonstrating that ASR-generated outputs deliver competitive SER performance relative to ground-truth transcripts.

Beyond transcripts, several studies have investigated ASR model representations directly for SER. Lu et al. \cite{lu2020speech} employed pre-trained features from end-to-end ASR models for SER, showing that hidden states from ASR encoders integrate both acoustic and linguistic information. Yeh et al. \cite{yeh2020speech} proposed a factorized adaptation approach that learns disentangled speaker and linguistic representations from end-to-end ASR models, achieving effective transfer to emotion recognition tasks. Li et al. \cite{li2022fusing} proposed the hierarchical attention fusion of audio features, ASR hidden states, and ASR transcripts, achieving SER performance comparable to that using ground-truth text. Lin and Wang \cite{lin2023robust} explored complementary semantic information from audio to mitigate ASR errors, employing an attention mechanism to calculate weighted acoustic representations on the basis of ASR hypotheses. Gao et al. \cite{gao2024speech} further designed a multi-level extraction framework that integrates acoustic and semantic information, where semantic representations are derived from ASR outputs and combined with acoustic features through cross-attention and gating. He et al. \cite{he2024mf} introduced two auxiliary tasks, ASR error detection and correction, to improve the semantic coherence of ASR text, and proposed a novel multimodal fusion method to learn shared representations across modalities.

However, most prior studies have focused on clean data conditions and overlooked an intrinsic property of ASR models: while encoders capture rich acoustic-linguistic cues, decoders also encode complementary semantic and contextual information through their language modeling role \cite{radford2023robust,sutskever2014sequence}. Such information is particularly valuable for emotion recognition, where lexical and prosodic cues jointly contribute to performance \cite{wagner2023dawn}. Meanwhile, the potential utility of decoder-derived representations has received little attention, and the relative robustness of encoder versus decoder features under diverse noise conditions has not been systematically examined. To address this gap, our work investigates both encoder and decoder representations as noise-robust features for NSER, providing the first comprehensive analysis of their effectiveness and complementarity in realistic acoustic environments.

\section{Proposed Method}
In this section, we describe in detail our proposed NSER model, which integrates a large-scale ASR model. As depicted in Fig. 1, the network comprises two primary components: an embedding module tasked with encoding noisy speech utilizing the ASR model and an emotion recognition module responsible for identifying the emotion label. These components will be further elaborated upon in subsequent sections.

\begin{figure}[htbp]
    \centering
    \includegraphics[width=\linewidth]{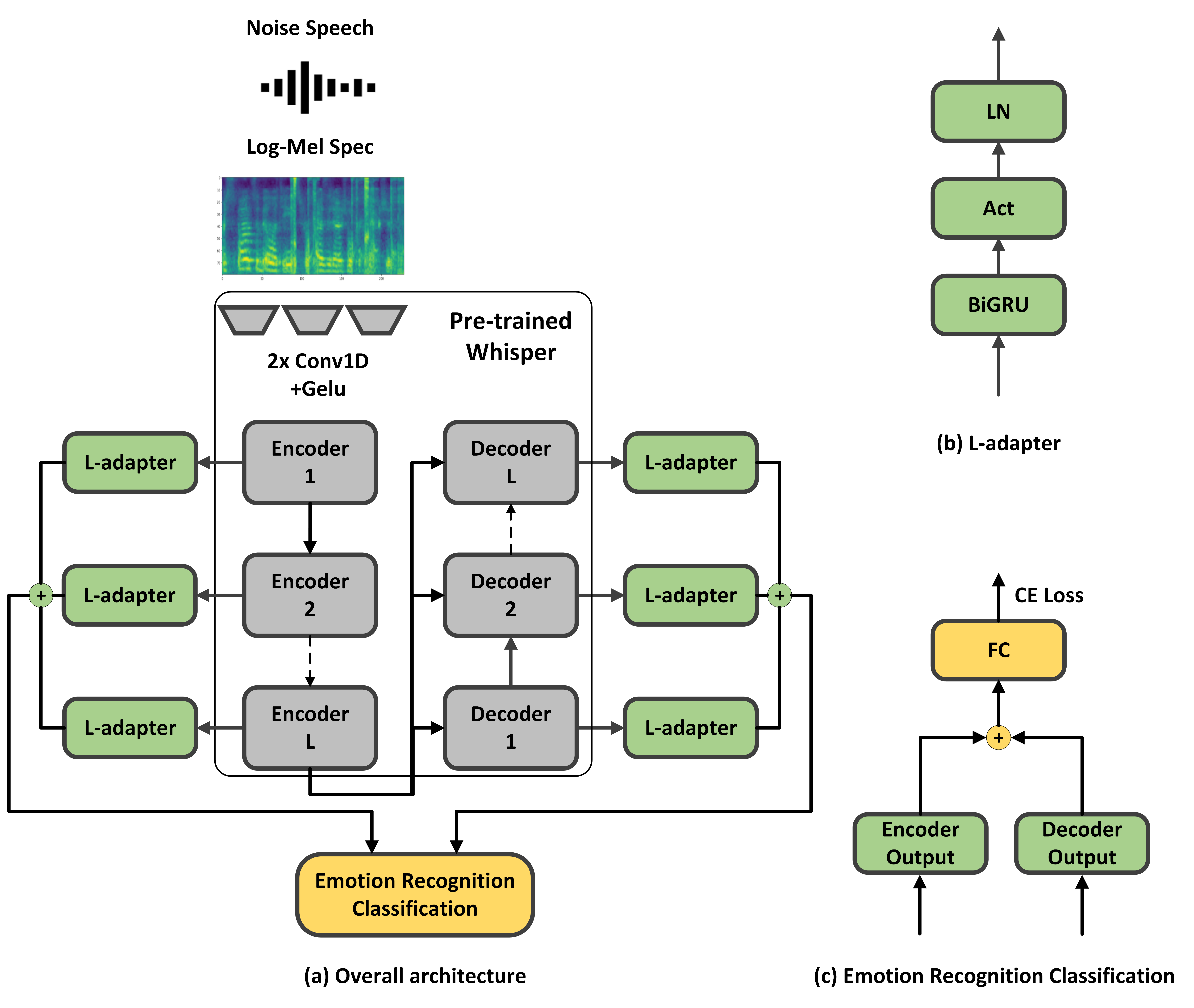}
    \caption{Structure of the NSER via ASR representations framework, where (a) represents the overall architecture, (b) details the structure of the L-adapter, and (c) outlines the structure of emotion recognition classification.}
\end{figure}

\subsection{Model Description}
As depicted, raw noisy audio utterances are fed into dedicated encoder and decoder networks designed to extract speech representations. The layer adapter module then processes these representations at various levels to extract multi-level emotional representations. These representations are subsequently input into the speech emotion classification model to obtain the emotion label.

\subsection{ASR Embedding Module}
Different from most existing models that recognize emotional states using SSL and text features, the proposed model utilizes ASR representations to offer advancements in emotion recognition from several perspectives. Whereas SSL presents the advantage of leveraging extensive unsupervised data for model training, recent ASR systems such as Whisper have primarily benefited from supervised training with substantial labeled datasets. Generally, when sufficient labeled data is available, supervised learning methods tend to exhibit superior performance. Additionally, ASR representations can capture more speech temporal structures than text features. Therefore, we opt for ASR as our feature extractor.

In this study, we use Whisper \cite{gong2023Whisper}, a robust supervised speech recognition model known for its noise robustness. Whisper is designed to transcribe spoken language into text and has been trained on a diverse dataset of approximately 680,000 hours of speech in 60 languages \cite{radford2023robust}. The model employs an encoder-decoder transformer architecture, where the encoder processes Mel-spectrogram inputs, and the decoder generates text sequences.

We define the mapping from the audio input \( x \) to the speech representation \( y \) using Whisper as follows:
\begin{equation}
y = F_\theta(x),
\end{equation}
where \( F_\theta \) denotes the mapping function implemented by Whisper. The audio input \( x \in \mathbb{R}^{t \times f} \) is a 2D feature matrix, where \( t \) represents the number of time steps and \( f \) denotes the feature dimension. The speech representation \( y \in \mathbb{R}^{m \times d} \) is a sequence of hidden representations produced by Whisper, where \( m \) is the sequence length and \( d \) is the feature dimension. 

The speech representation \( y \) undergoes initial preprocessing through a convolutional neural network (CNN) layer to obtain the preliminary representation \( H_0 \):
\begin{equation}
H_0 = \mathrm{CNN}(y),
\end{equation}

Subsequently, the initial hidden states \( H_0 \) are processed through \( L \) encoder layers. Each layer updates its hidden states as follows:
\begin{equation}
H_{l+1} = \mathrm{Encoder}(H_l) \quad \text{for} \quad l = 1, \ldots, L-1,
\end{equation}
where \( m \) is the length of the input sequence (i.e., the number of time steps or frames in the input), and \( H_l = (h_l^1, h_l^2, \ldots, h_l^m) \in \mathbb{R}^{m \times d} \) represents the hidden states at layer \( l \). The final encoder output, \( H_L \), encapsulates the contextual information of the entire input sequence, preparing it for the decoding process.

For the decoder, the initialization begins with a sequence of discrete tokens, starting with a special start-of-sequence token. During inference, the decoder generates tokens autoregressively, where each newly generated token is appended to the input sequence for the next decoding step. At each step, the decoder conditions the previously generated tokens and the contextual information \( H_L \) from the encoder to produce the next token:
\begin{equation}
D_{l+1} = \mathrm{Decoder}(D_l, H_L) \quad \text{for} \quad l = 0, 1, \ldots, L-1,
\end{equation}
where \( D_l \) represents the decoder’s hidden states at step \( l \), and \( n \) corresponds to the length of the generated output sequence. Once the decoding process is complete, the final hidden states \( D_L = (d_L^1, d_L^2, \ldots, d_L^n) \in \mathbb{R}^{n \times d} \) are passed through a fully connected layer and a softmax activation to produce the probability distribution over the vocabulary.

\subsection{Layer Adapter Module (L-adapter)}
To leverage intermediate representations from the initial fine-tuning stages, layer adapters create pathways from each of the encoder and decoder layers to the emotion recognition classification module. Each layer adapter consists of a bi-directional gated recurrent unit (BiGRU) layer, followed by 1D max pooling, a non-linear ReLU activation function, and a normalization layer. The BiGRU network is an advanced variant of the RNN designed to address the vanishing gradient problem \cite{cho2014learning}, which is common in sequential data analysis. It has been widely used in emotion-related tasks \cite{shi23e_interspeech,shi2024multimodal,li2023music}. Additionally, following a previous study \cite{li2022fusing}, as illustrated in Fig 1(b):
\begin{equation}
\alpha^{(E)}_l = \mbox{Maxpooling}(\mathrm{BiGRU}(H_l))
\end{equation}
for \( l = 1, 2, \ldots, L \) in the encoder, and

\begin{equation}
\alpha^{(D)}_l = \mbox{Maxpooling}(\mathrm{BiGRU}(D_l))
\end{equation}
for \( l = 1, 2, \ldots, L \) in the decoder.

The weighted sum of the adapted representations is computed as
\begin{equation}
h^\ast = \sum_{l=1}^{L} w^{(E)}_l \alpha^{(E)}_l + \sum_{l=1}^{L} w^{(D)}_l \alpha^{(D)}_l.
\end{equation}
This is fed into the emotion recognition classification, where \( w^{(E)}_l \) and \( w^{(D)}_l \) are learnable weights.

\subsection{Emotion Recognition Classification}
As shown in Fig. 1(c), we use a fully connected layer as the classifier for emotion recognition.
\begin{equation}
P(y_{\rm emo} \mid x) = \mathrm{SoftMax}(\mathrm{FC}(h^\ast))
\end{equation}

The loss function \( \mathcal{L} \) used for final emotion classification computes the softmax cross-entropy between the predicted probabilities \( P(y_{\rm emo} \mid x) \) and the ground truth labels \( y_{\rm emo} \):
\begin{equation}
\mathcal{L} = - \sum_{i} y_{\rm emo}^{(i)} \log P(y_{\rm emo}^{(i)} \mid x),
\end{equation}
where \( P(y_{\rm emo}^i \mid x) \) represents the predicted probability distribution over emotion classes given the speech \( x \), and \( y_{\rm emo}^i \) denotes the ground truth label for the \( i \)-th sample.

\section{Experimental settings}
\subsection{Database}
In this study, we employ three emotional speech corpora. The MELD dataset, which contains mixed recordings from realistic acoustic environments, has been widely used for NSER research \cite{wu2023metricaug, kshirsagar2023task}. The IEMOCAP dataset provides relatively clean and controlled recordings, and the CASIA dataset is used to explore cross-lingual scenarios. The emotional distribution across these datasets is illustrated in Fig. 2.

\label{sec:format}
\begin{figure}[htbp]
    \centering
    \includegraphics[width=\linewidth]{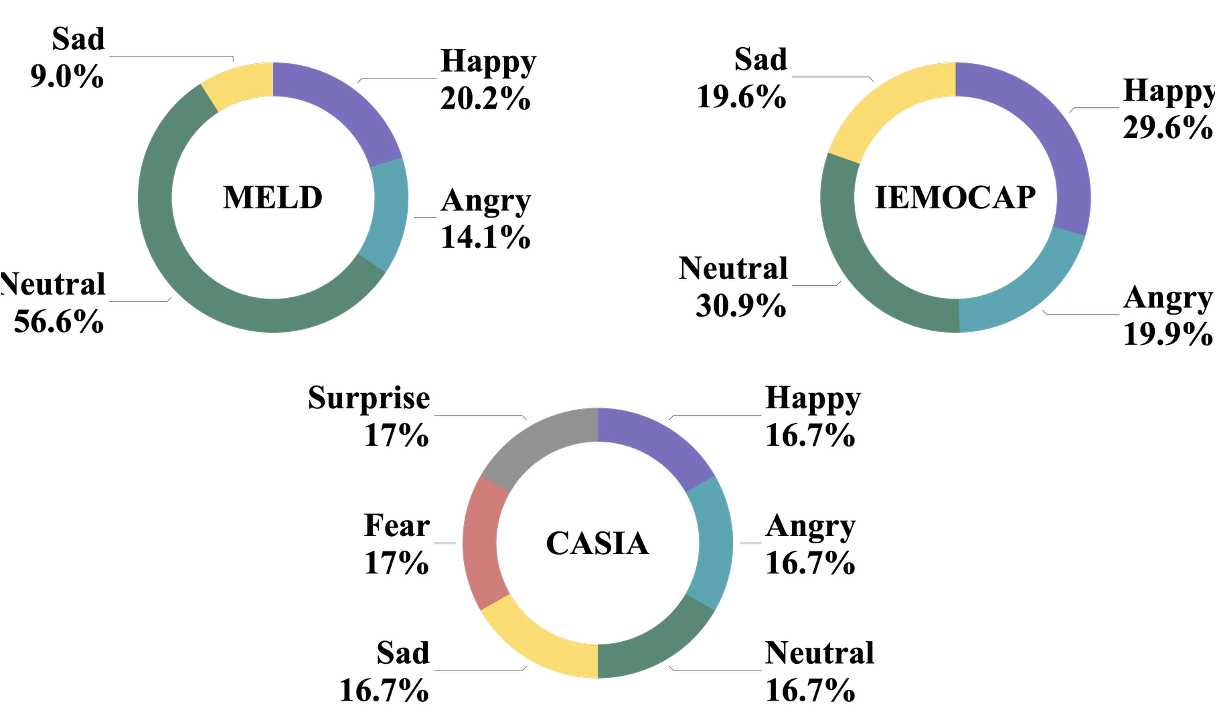}
    \caption{Amount of emotion distribution in MELD, IEMOCAP, and CASIA.}
\end{figure}

\subsubsection*{\bf 1) MELD}
The MELD dataset is an extension of a sentiment-focused dataset, comprising 1,433 dialogues from the American drama “Old Friends,” totaling 13,708 sentences and incorporating various data types, including video, text, and speech \cite{poria2018meld}. Various environmental noises are recorded within the MELD dataset, such as car horn sounds, knocking noises, plate rattling sounds, and more. Each dialogue in MELD is annotated with seven emotions: anger, disgust, sadness, happiness, neutrality, surprise, and fear. Additionally, the dataset provides standard training, validation, and test sets to facilitate comparative experimental results. Our study focused on four emotions: happiness, anger, sadness, and neutrality.

\subsubsection*{\bf 2) IEMOCAP}	
The Interactive Emotional Dyadic Motion Capture (IEMOCAP) dataset is widely used in affective computing \cite{busso2008iemocap}. It consists of approximately 12 hours of audio-visual recordings designed specifically for two-person dialogues. Each dialogue is segmented into utterances and annotated with continuous labels in the Valence-Arousal dimension, along with categorical labels for emotional states.

For this study, we focus on four categorical emotions: neutral, happiness, sadness, and anger, aligning with experimental protocols used in previous research \cite{li2023music,gao2023two}. Additionally, we combine happiness and excitement into a single category. Given that the IEMOCAP recordings were primarily captured in clean environments, we introduce noise from the DEMAND corpus to simulate realistic acoustic conditions. The DEMAND corpus includes six noise categories: Street, Domestic, Office, Public, Nature, and Transportation \cite{thiemann2013diverse}. For our experiments, we selected Office (representing indoor environments), Nature, and Transportation (representing outdoor environments) to construct distinct noisy scenarios.

\subsubsection*{\bf 2.1 IEMOCAP-Human}
To replicate human speech noise in realistic acoustic environments, we merged the IEMOCAP data with the Office category from the DEMAND corpus. We used the OMEETING segment of the Office data, recorded in a busy cafeteria, to augment the IEMOCAP dataset.

\subsubsection*{\bf 2.2 IEMOCAP-Environment (Park)}
To simulate natural noise in realistic acoustic environments, we combined the IEMOCAP data with the Nature category from the DEMAND corpus. Specifically, we used the NPARK segment of the Nature data, recorded in a bustling park, to augment the IEMOCAP dataset.

\subsubsection*{\bf 2.3 IEMOCAP-Environment (Traffic)}
To simulate traffic noise in realistic acoustic environments, we combined the IEMOCAP data with the Transportation category from the DEMAND corpus. We used the Traffic segment of the Transportation data, recorded on a busy roadway, to augment the IEMOCAP dataset.

\subsubsection*{\bf 3) CASIA}	
The CASIA Mandarin Chinese Emotional Corpus was meticulously designed in 2005, featuring both parallel and non-parallel transcript recordings conducted in a professional studio \cite{zhang2008design}. The set includes 100 parallel sentences intended to mitigate the effect of textual content on emotion recognition. In this study, we leveraged a parallel transcript section containing six emotional states from four speakers: anger, fear, happiness, neutrality, sadness, and surprise, amounting to 7200 utterances. 

\subsection{Experimental Procedure}
In this study, we conducted seven experiments to investigate various aspects of ASR representations in NSER.

In Experiment 1, we evaluated a comprehensive set of baselines for NSER under 0 dB SNR across the MELD, IEMOCAP, IEMOCAP-Human, and IEMOCAP-Environment (Park, Traffic) datasets. As a starting point, we considered the Mel-spectrogram representation as a conventional NSER baseline. We further extracted HSF features using the OpenSMILE toolbox\footnote{\url{https://github.com/audeering/opensmile}}, which provides extensive acoustic feature configurations, including IS09\_emotion, IS10\_paraling, IS11\_speaker\_state, IS12\_speaker\_trait, IS13\_ComParE, and eGeMAPS. For noise reduction, we adopted Conv-TasNet for speech separation and DCCRN for speech enhancement. We additionally evaluated mainstream NSER models such as MLDK and MV-BiSSM. In terms of self-supervised learning (SSL) methods, we fine-tuned \textit{Wav2Vec 2.0}\footnote{\url{https://huggingface.co/facebook/wav2vec2-base}}, \textit{HuBERT}\footnote{\url{https://huggingface.co/facebook/hubert-base-ls960}}, and \textit{WavLM}\footnote{\url{https://huggingface.co/microsoft/wavlm-base}} on noisy speech. Finally, we compared ASR-based representations, including a Conformer-based ASR model as the conventional baseline \cite{gulati2020conformer, ta2022improving, seo2022multi}, as well as large-scale ASR models such as SpeechLM \cite{zhang2024speechlm} and Whisper, all integrated with the same adapter mechanism on both encoder and decoder layers as in our proposed method. A summary of all baselines is provided in Table I.

\begin{table*}[htbp]
\centering
\small
\renewcommand{\arraystretch}{1.2} % 调整行距
\caption{Summary of baselines used in Experiment 1.}
\begin{tabularx}{\textwidth}{c|c|X}
\toprule
\textbf{Method} & \textbf{Model} & \textbf{Description} \\
\midrule
\midrule
-- & BiGRU & Traditional NSER baseline using Mel-Spectrogram. \\
\hline
\multirow{6}{*}{HSF} 
& IS09\_emotion \cite{Schuller:2009} & 384 features; statistical functionals of basic Low-Level Descriptors (LLDs, e.g., energy, pitch). \\
& IS10\_paraling \cite{Schuller:2010} & 1582 features; 34 LLDs + deltas with functionals, plus pitch features. \\
& IS11\_speaker\_state \cite{Schuller:2011} & 4368 features; multiple LLD groups with utterance-level functionals. \\
& IS12\_speaker\_trait \cite{Schuller:2012} & 6125 features; extends IS11 with local-extrema functionals. \\
& IS13\_ComParE \cite{Schuller:2013} & 6373 features; comprehensive set of energy, spectral, cepstral, and voicing descriptors. \\
& eGeMAPS \cite{Eyben:2015} & 88 features; compact, theory-driven set with MFCCs, spectral flux, and voiced/unvoiced statistics. \\
\hline
\multirow{2}{*}{Denoising} 
& Conv-TasNet \cite{luo2019conv} & Convolutional time-domain network for speech separation. \\
& DCCRN \cite{hu2020dccrn} & Deep complex convolutional recurrent network for speech enhancement. \\
\hline
\multirow{2}{*}{NSER} 
& MLDK \cite{liu2023multi} & Multi-level deep kernel learning model for NSER. \\
& MV-BiSSM \cite{liu2025enhanced} & Multi-view bidirectional sequential self-attention model for NSER. \\
\hline
\multirow{3}{*}{SSL} 
& Wav2Vec 2.0 \cite{baevski2020wav2vec} & Self-supervised model for learning contextualized speech representations. \\
& HuBERT \cite{hsu2021hubert} & Self-supervised model for learning robust speech embeddings through hidden-unit clustering. \\
& WavLM \cite{chen2022wavlm} & Self-supervised model for learning speaker- and environment-aware speech representations. \\
\hline
\multirow{6}{*}{ASR} 
& \multirow{2}{*}{Conformer \cite{gulati2020conformer}} & End-to-end ASR backbone combining convolution and transformer layers to effectively capture both local dependencies (e.g., phonetic details) and global context for speech recognition. \\
& \multirow{2}{*}{SpeechLM \cite{zhang2024speechlm}} & Multilingual speech–text pre-trained ASR model that unifies acoustic and linguistic information through joint objectives, enabling stronger transfer across languages and tasks. \\
& \multirow{2}{*}{Whisper \cite{radford2023robust}} & Large-scale encoder–decoder ASR model designed for robust multilingual and multitask speech recognition, covering diverse languages, accents, and noisy acoustic conditions. \\
\bottomrule
\end{tabularx}
\vspace{-5mm}
\end{table*}

In Experiment 2, we examined the effectiveness of different ASR representations for NSER using the MELD, IEMOCAP, IEMOCAP-Human, and IEMOCAP-Environment (Park, Traffic) datasets under 0 dB SNR conditions. Specifically, we compared: (i) single-layer representations (either the last layer or the best-performing layer among all layers), and (ii) fusion-based multi-layer representations, which were implemented through three strategies: simple fusion (mean pooling across layers), parameterized fusion (scalar mixing and weighted sum), and advanced fusion (attention-based pooling). In addition, we proposed the Layer Adapter approach, which extends weighted-sum fusion of ASR encoder and decoder representations by enabling more effective cross-layer integration, thereby producing richer and more robust features for NSER.

In Experiment 3, we examined the layer-wise effectiveness of ASR representations for NSER. Specifically, we compared representations extracted from the encoder and decoder layers of ASR models, whereas for SSL models only encoder-layer representations were considered. All experiments were conducted on the IEMOCAP, IEMOCAP-Human, and IEMOCAP-Environment (Park, Traffic) datasets at an SNR of 0 dB.

In Experiment 4, we examined the robustness of ASR representations for NSER under varying noise intensities. We conducted a comparative analysis using the IEMOCAP-Human and IEMOCAP-Environment (Park, Traffic) datasets, systematically controlling noise at SNR levels of -10, -5, 0, 5, and 10 dB. We evaluated different ASR representations, including: Whisper encoder with the Layer Adapter; Whisper decoder with the Layer Adapter; Combined Whisper encoder and decoder with the Layer Adapter, alongside representations from SSL models.

In Experiment 5, we investigated the correlation between ASR performance and NSER outcomes across different noise intensities. The comparative analysis was conducted using the IEMOCAP-Human dataset and the IEMOCAP-Environment dataset (Park, Traffic). For evaluation, we considered ASR representations from two models: a Conformer-based ASR model as a conventional baseline, and Whisper, a large-scale ASR model.

In Experiment 6, we evaluated the effectiveness of ASR representations across different modalities for NSER, using the MELD, IEMOCAP, IEMOCAP-Human, and IEMOCAP-Environment (Park, Traffic) datasets under 0 dB SNR. We conducted a comparative analysis between speech- and text-based modalities, leveraging SSL methods on both ASR-generated and ground-truth transcriptions. For the text modality, we employed a BERT model, following prior studies on emotion recognition \cite{adoma2020comparative,kumar2022bert}. For the speech modality, we compared ASR representations from Whisper, including the encoder with an adapter, the decoder with an adapter, and a combined encoder–decoder configuration with an adapter.

In Experiment 7, we examined the impact of cross-lingual settings on the performance of ASR representations in NSER. Although ASR systems effectively captured speech and suppressed background noise, their performance degraded when handling languages not included in their training data. Using the CASIA dataset (Mandarin Chinese) as a monolingual scenario, we compared SSL representations under cross-lingual conditions with ASR representations under both monolingual and cross-lingual conditions. For SSL, we evaluated \textit{wav2vec 2.0-base}\footnotemark[1], \textit{HuBERT-base}\footnotemark[2], and \textit{WavLM-base}\footnotemark[3], pre-trained on 960 hours of English Librispeech data, in frozen and fine-tuned configurations. For ASR, we employed two Whisper variants: \textit{Whisper-CN}\footnote{\url{https://huggingface.co/Jingmiao/whisper-small-chinese_base}}, trained solely on Mandarin data to represent monolingual conditions, and \textit{Whisper-EN}\footnote{\url{https://huggingface.co/openai/whisper-small.en}}, trained on English data to represent cross-lingual conditions. Moreover, since CASIA provides parallel utterances with consistent textual content across different emotions, this experiment mitigates the confounding influence of linguistic information. Consequently, the results more faithfully reflect the contributions of paralinguistic cues captured by ASR representations in NSER.

\vspace{-2mm}
\subsection{Implementation}
All models were implemented in Python 3.7 with PyTorch 1.11.0. Training and evaluation were conducted on a server equipped with an Intel(R) Xeon(R) Gold 6248 CPU @ 2.50 GHz, 32 GB RAM, and a single NVIDIA Tesla V100 GPU.

In the speech modality, we employed \textit{Conformer-small}\footnote{\url{https://huggingface.co/speechbrain/asr-conformer-transformerlm-librispeech}}
 as a representative conventional ASR model, and additionally considered two large-scale approaches: (i) \textit{Whisper-small}\footnotemark[4] and (ii) \textit{SpeechLM-Base}\footnote{\url{https://github.com/microsoft/SpeechT5/tree/main/SpeechLM}}
, both adopting a transformer encoder–decoder architecture with 12 encoder layers, 12 decoder layers, and a hidden size of 768. For the text modality, we used \textit{BERT-base}\footnote{\url{https://huggingface.co/google-bert/bert-base-uncased}}
, featuring a hidden size of 768, 12 attention layers, and 12 heads.

For baseline experiments, we implemented several comparison models. For Mel-spectrogram and denoised Mel-spectrogram features, a BiGRU with 256 units and dropout 0.5 was employed. For HSF features, we adopted a fully connected network with 512 units and dropout 0.5. For SSL-based methods, mean pooling was applied along the temporal dimension, followed by a fully connected classifier. To ensure fairness in model capacity, all SSL models were of Base size (hidden size 768). Detailed settings are shown in Table~II.

\begin{table}[htbp]
\centering
\caption{Hyperparameters (top) and model configurations (bottom) for baseline models, including Mel-spectrogram and denoised Mel-spectrogram (Mel/Denoising), HSF, and SSL.}

\label{tab:params}
\footnotesize
\renewcommand{\arraystretch}{1.2} % 调整行距
\resizebox{\linewidth}{!}{
\begin{tabular}{lccc}
\toprule
\textbf{Baseline} & \textbf{Mel / Denoising} & \textbf{HSF} & \textbf{SSL} \\
\midrule
\midrule
\multicolumn{4}{l}{\textit{Hyperparameters}} \\
\midrule
\midrule
Batch size        & 32            & 32            & 32 \\
Epochs            & 50            & 50            & 50 \\
Learning rate     & $1 \times 10^{-4}$ & $1 \times 10^{-4}$ & $1 \times 10^{-4}$ \\
Dropout           & 0.5           & 0.5           & 0.5 \\
Optimizer         & Adam          & Adam          & Adam \\
Loss              & Cross-Entropy & Cross-Entropy & Cross-Entropy \\
\midrule
\midrule
\multicolumn{4}{l}{\textit{Model configurations}} \\
\midrule
\midrule
Backbone frozen   & No            & No            & No \\
Hidden size       & 256           & 512           & 768 \\
Downstream head   & BiGRU         & FC            & MeanPool + FC \\
\bottomrule
\end{tabular}}
\end{table}

For ASR-based models, consistent with prior work, the Whisper, Conformer, and SpeechLM backbones were frozen during training. The proposed Layer Adapter employed a BiGRU with 256 units per direction and a dropout rate of 0.5, with the embedding size fixed at 512 across adapters. The emotion classifier was a fully connected layer with a dropout of 0.5. Training used a batch size of 32, the Adam optimizer with a learning rate of $1 \times 10^{-4}$, and cross-entropy loss.

\subsection{Evaluation Metrics}
In evaluating our results on IEMOCAP, where a standard train, dev, and test split is not predefined, we employed a five-fold cross-validation approach, consistent with previous studies \cite{sun2024fine,shi23e_interspeech}. For MELD, following established practices in prior research \cite{ghosal2020cosmic, shi23e_interspeech}, we utilized the Train and Validation sets for training, with the Test set used for evaluation. As for CASIA, we adopted a four-fold cross-validation approach, aligning with previous work \cite{peng2017speech}.

For NSER performance evaluation, we used unweighted average recall (UAR) and F1 scores, which are widely recognized metrics for assessing models on imbalanced datasets \cite{chen20183}. These metrics were applied across four distinct emotional categories. To evaluate the ASR performance, we used word error rate (WER), which is commonly employed to gauge recognition accuracy in speech-related tasks \cite{ten2003emotions}.

\section{Results and Evaluation}
\subsection{Effect of ASR Representations on NSER}	
In this study, we conducted a comprehensive comparative analysis to assess the efficacy of ASR representations in NSER, benchmarking them against traditional HSF features, noise reduction approaches, mainstream NSER models, and SSL methods. The results are presented in Table III.

\begin{table*}[htbp]
\caption{Comparison of NSER performance across different feature representations, including noisy Mel-Spectrograms, HSF features, denoising methods, NSER baselines, SSL representations (\textbf{FT}: fine-tuning), and ASR representations (\textbf{FZ}: frozen). Results are reported on MELD, IEMOCAP, and IEMOCAP under Human and Environment (Park/Traffic) noise.}
\centering
\renewcommand{\arraystretch}{1.5}
\resizebox{\linewidth}{!}{
\begin{tabular}{c|c|cc|cc|cc|cc|cc}
\toprule
\multirow{3}{*}{\textbf{Method}}
  & \multirow{3}{*}{\textbf{Feature}}
  & \multicolumn{2}{c|}{\multirow{2}{*}{\textbf{MELD}}}
  & \multicolumn{2}{c|}{\multirow{2}{*}{\textbf{IEMOCAP}}}
  & \multicolumn{2}{c|}{\multirow{2}{*}{\makecell{\textbf{IEMOCAP}\\\textbf{(Human)}}}}
  & \multicolumn{4}{c}{\makecell{\textbf{IEMOCAP}\\\textbf{(Environment)}}} \\
\cline{9-12}
& & \multicolumn{2}{c|}{} & \multicolumn{2}{c|}{} & \multicolumn{2}{c|}{}
  & \multicolumn{2}{c|}{\textbf{Park}} & \multicolumn{2}{c}{\textbf{Traffic}} \\
\cline{3-12}
& & \textbf{UAR(\%)} & \textbf{F1(\%)} & \textbf{UAR(\%)} & \textbf{F1(\%)} & \textbf{UAR(\%)} & \textbf{F1(\%)} & \textbf{UAR(\%)} & \textbf{F1(\%)} & \textbf{UAR(\%)} & \textbf{F1(\%)} \\
\midrule
\midrule
 -- & Mel-Spectrogram & 32.73 & 33.41 & 57.12 & 56.10 & 53.22 & 52.61 & 53.71 & 53.34 & 54.00 & 52.83 \\
\hline
\multirow{6}{*}{HSF}
  & IS09\_emotion & 31.92 & 31.28 & 58.65 & 57.84 & 56.31 & 55.43 & 54.73 & 54.33 & 54.81 & 53.87 \\
  & IS10\_paraling & 34.90 & 35.34 & 61.22 & 59.49 & 56.34 & 55.96 & 56.15 & 55.01 & 56.31 & 55.43 \\
  & IS11\_speaker\_state & 37.06 & 35.88 & 60.17 & 59.83 & 55.12 & 54.06 & 55.57 & 53.96 & 55.12 & 54.06 \\
  & IS12\_speaker\_trait & 35.22 & 35.10 & 60.46 & 59.11 & 55.31 & 53.56 & 54.38 & 52.80 & 55.31 & 53.56 \\
  & IS13\_ComParE & 37.33 & 36.04 & 60.11 & 59.30 & 54.84 & 53.31 & 55.85 & 54.42 & 54.84 & 53.31 \\
  & eGeMAPS & 34.04 & 33.36 & 58.79 & 58.02 & 53.47 & 52.46 & 54.75 & 54.02 & 53.95 & 53.41 \\
\hline
\multirow{2}{*}{Denoising}
  & Conv-TasNet    & 33.16 & 33.44 &   -   &   -   & 52.21 & 51.68 & 55.03 & 53.84 & 55.62 & 54.17 \\
  & DCCRN          & 33.39 & 33.24 &   -   &   -   & 51.61 & 51.30 & 53.84 & 52.82 & 55.12 & 53.48 \\
\hline
\multirow{2}{*}{NSER}
  & MLDK & 45.51 & 45.25 & 70.65 & 69.11 & 65.37 & 63.58 & 68.82 & 67.29 & 70.21 & 68.88 \\
  & MV-BiSSM & 47.71 & 47.45 & 75.89 & 74.30 & 68.51 & 66.74 & 72.66 & 72.01 & 73.97 & 73.52 \\
\hline
\multirow{3}{*}{SSL}
  & Wav2Vec 2.0 (FT) & 43.00 & 43.46 & 68.08 & 67.19 & 59.24 & 57.49 & 60.68 & 59.52 & 64.12 & 62.63 \\
  & HuBERT (FT) & 42.59 & 42.61 & 66.91 & 66.32 & 61.08 & 60.08 & 61.41 & 60.72 & 62.66 & 62.04 \\
  & WavLM (FT) & 40.29 & 41.05 & 65.29 & 64.22 & 59.28 & 57.36 & 60.97 & 58.74 & 62.28 & 60.95 \\
\hline
\multirow{3}{*}{ASR}
  & Conformer (FZ) & 39.97 & 40.27 & 72.84 & 73.04 & 66.18 & 65.86 & 68.94 & 69.17 & 70.54 & 70.96 \\
  & SpeechLM (FZ) & 44.27 & 45.68 & 73.18 & 74.79 & 66.59 & 66.74 & 69.06 & 70.49 & 71.71 & 71.14  \\
\cline{2-12}
  & \textbf{Proposed}
                   & \textbf{48.95} & \textbf{48.87}
                   & \textbf{76.50} & \textbf{76.53}
                   & \textbf{68.79} & \textbf{68.39}
                   & \textbf{73.00} & \textbf{72.42}
                   & \textbf{74.39} & \textbf{74.04} \\
\bottomrule
\end{tabular}}
\end{table*}

The results highlight the superior performance of ASR representations across various datasets. On the MELD dataset, the proposed method achieves the largest gains. Compared with noisy Mel-Spectrograms, it improves by 16.22\% in UAR and 15.46\% in F1. Against the best HSF features (IS13), the gains are 11.62\% and 12.83\%. Over denoising baselines, the proposed method achieves 15.56\% improvement in UAR and 15.63\% in F1 compared with DCCRN. Relative to the strongest NSER baseline (MV-BiSSM), it adds 1.24\% in UAR and 1.42\% in F1. Compared with Wav2Vec 2.0, the best SSL representation, the proposed approach achieves 5.95\% and 5.41\% improvements. Finally, even against SpeechLM, a large-scale ASR model, the proposed model still surpasses by 4.68\% in UAR and 3.19\% in F1.

On the IEMOCAP dataset, the proposed approach improves by 19.38\% in UAR and 20.43\% in F1 over noisy Mel-Spectrograms. Relative to the HSF feature (IS10), it gains 15.28\% in UAR and 17.04\% in F1. Against MV-BiSSM, the strongest NSER baseline, the proposed method provides 0.61\% and 2.23\% improvements. Compared with Wav2Vec 2.0, the best SSL, the gains are 8.42\% and 9.34\%. Over SpeechLM, it improves by 3.32\% in UAR and 1.74\% in F1.

For the IEMOCAP-Human subset, the proposed method shows 15.57\% and 15.78\% improvements over noisy Mel-Spectrograms. Compared with the best HSF features (IS10), the gains reach 12.45\% in UAR and 12.43\% in F1. Over Conv-TasNet, it improves by 16.58\% and 16.71\%. Compared with MV-BiSSM, the proposed method is slightly higher, achieving 0.28\% and 1.65\% gains. Against HuBERT, the best SSL, it achieves 7.71\% and 8.31\% gains, and over SpeechLM it further improves by 2.20\% and 1.65\%.

For the IEMOCAP-Environment subset (Park), the proposed method improves by 19.29\% and 19.08\% compared with noisy Mel-Spectrograms. Over the best HSF features (IS10), it achieves 16.85\% and 17.41\% improvements. Against ConvTasNet, the gains are 17.97\% and 18.58\%. Compared with MV-BiSSM, the proposed method is slightly higher, with improvements of 0.34\% in UAR and 0.41\% in F1. Compared with HuBERT, the best SSL, it improves by 11.59\% and 11.70\%. Over SpeechLM, the gains are 3.94\% in UAR and 1.93\% in F1.

For the IEMOCAP-Environment subset (Traffic), the proposed method shows improvements of 20.39\% and 21.21\% over noisy Mel-Spectrograms. Compared with the best HSF features (IS10), the gains are 18.08\% in UAR and 18.61\% in F1. Over Conv-TasNet, it improves by 18.77\% and 19.87\%. Relative to MV-BiSSM, the proposed method improvements of 0.42\% in UAR and 0.52\% in F1. Compared with Wav2Vec 2.0, the best SSL model, it achieves 10.27\% and 11.41\% improvements. Over SpeechLM, the gains are 2.68\% in UAR and 2.90\% in F1.

In summary, the proposed ASR-based approach, exemplified by the Whisper encoder–decoder framework, achieves consistently strong performance over HSF features, conventional denoising, prior NSER baselines, and SSL models, while demonstrating notable robustness to environmental and human noise through joint use of encoder and decoder representations.

\vspace{-2mm}
\subsection{Effect of Different ASR Representations}	

\begin{table*}[htbp]
\caption{Comparison of NSER performance was conducted using ASR representation at encoder and decoder levels.}
\centering
\renewcommand{\arraystretch}{1.5}
\resizebox{\linewidth}{!}{
\begin{tabular}{c|c|cc|cc|cc|cc|cc} 
\toprule
\multirow{3}{*}{\textbf{Level}} & \multirow{3}{*}{\textbf{Method}} 
  & \multicolumn{2}{c|}{\multirow{2}{*}{\textbf{MELD}}}
  & \multicolumn{2}{c|}{\multirow{2}{*}{\textbf{IEMOCAP}}}
  & \multicolumn{2}{c|}{\multirow{2}{*}{\makecell{\textbf{IEMOCAP}\\\textbf{(Human)}}}}
  & \multicolumn{4}{c}{\makecell{\textbf{IEMOCAP}\\\textbf{(Environment)}}} \\
\cline{9-12}
&   & \multicolumn{2}{c|}{} & \multicolumn{2}{c|}{} & \multicolumn{2}{c|}{}
    & \multicolumn{2}{c|}{\textbf{Park}} & \multicolumn{2}{c}{\textbf{Traffic}} \\
\cline{3-12}
& & \textbf{UAR(\%)} & \textbf{F1(\%)} & \textbf{UAR(\%)} & \textbf{F1(\%)} & \textbf{UAR(\%)} & \textbf{F1(\%)} & \textbf{UAR(\%)} & \textbf{F1(\%)} & \textbf{UAR(\%)} & \textbf{F1(\%)} \\
\midrule
\midrule
\multirow{6}{*}{Encoder} 
  & Last             & 44.56 & 45.69 & 72.71 & 72.11 & 61.96 & 61.64 & 68.05 & 67.30 & 69.86 & 69.93\\
  & Best             & 44.56 & 45.69 & 72.71 & 72.11 & 63.44 & 62.61 & 68.05 & 67.30 & 69.86 & 69.93\\
\cline{2-12}
  & Mean & 46.36 & 46.46 & 72.91 & 72.19 & 65.38 & 65.21 & 68.54 & 67.83 & 69.61 & 69.01  \\
  & Attention & 46.65 & 46.87  & 73.39 & 73.22 & 65.70  & 65.43 & 69.51 & 69.02 & 70.74 & 70.52 \\
  & Scalar Mixing & 46.69 & 47.88 & 74.25 & 74.06 & 65.94 & 65.58 & 69.87 & 69.24 & 71.64 & 71.34 \\
  & Weighted Sum & 46.71 & 48.01 & 75.02 & 74.10 & 67.04 & 66.36 & 70.56 & 69.43 & 72.21 & 71.87 \\
\hline
\multirow{6}{*}{Decoder} 
  & Last             & 45.55 & 46.55 & 70.27 & 67.91 & 60.11 & 58.81 & 65.22 & 63.43 & 66.05 & 63.91 \\
  & Best             & 46.05 & 47.35 & 71.73 & 71.40 & 62.70 & 62.02 & 67.74 & 67.47 & 68.59 & 67.54\\
\cline{2-12}
  & Mean & 45.00 & 46.58 & 70.70 & 70.29 & 62.07 & 61.42 & 65.92 & 65.62 & 66.11 & 65.93 \\
  & Attention & 46.21 & 46.98  & 71.03 & 70.15 & 62.91 & 62.23 & 67.27& 66.77 & 69.91 & 69.09 \\
  & Scalar Mixing & 46.52 & 47.46 & 71.74 & 72.17 & 64.33 & 64.34 & 69.23 & 69.47 & 71.42 & 70.52 \\
  & Weighted Sum & 46.37 & 47.55 & 72.75 & 72.39 
                      & 66.99 & 66.61 & 69.72 & 68.39 & 71.90 & 69.12\\
\hline
ALL 
  & \textbf{Proposed}        & \textbf{48.95} & \textbf{48.87} & \textbf{76.50} & \textbf{76.53} 
                      & \textbf{68.79} & \textbf{68.39} & \textbf{73.00} & \textbf{72.42} & \textbf{74.39} & \textbf{74.04}\\
\bottomrule
\end{tabular}}
\end{table*}

In Experiment 2, we evaluated the impact of ASR representations for NSER under 0 dB SNR across MELD, IEMOCAP, IEMOCAP-Human, and IEMOCAP-Environment (Park, Traffic). The representations were grouped into Single-layer (last, best-performing layer) and All-layer (Mean, Attention, Scalar mixing, Weighted sum), and compared with the proposed Layer Adapter method, as detailed in Table IV.

For the MELD dataset, the best Single-layer representation (Decoder, best-performing layer) achieved 46.05\% in UAR and 47.35\% in F1. Among Multi-layer fusion methods, the Encoder-based Weighted Sum reached 46.71\% in UAR and 48.01\% in F1. In comparison, the proposed Layer Adapter achieved 48.95\% in UAR and 48.87\% in F1, yielding improvements of 2.90\% in UAR and 1.52\% in F1 over the best Single-layer, and 2.24\% in UAR and 0.86\% in F1 over the best Multi-layer representation.

For the IEMOCAP dataset, the best Single-layer representation (Encoder, best-performing layer) achieved 72.71\% in UAR and 72.11\% in F1. Among Multi-layer fusion methods, the Encoder-based Weighted Sum reached 75.02\% in UAR and 74.10\% in F1. In comparison, the proposed Layer Adapter attained 76.50\% in UAR and 76.53\% in F1, corresponding to improvements of 3.79\% in UAR and 4.42\% in F1 over the best Single-layer, and 1.48\% in UAR and 2.43\% in F1 relative to the best Multi-layer representation.

For the IEMOCAP-Human subset, the best Single-layer representation (Encoder, best-performing layer) achieved 63.44\% in UAR and 62.61\% in F1. The best Multi-layer fusion method (Decoder, Weighted Sum) attained 66.99\% in UAR and 66.61\% in F1. In contrast, the proposed method delivered 68.79\% in UAR and 68.39\% in F1, giving improvements of 5.35\% in UAR and 5.78\% in F1 over the best Single-layer, and 1.80\% in UAR and 1.78\% in F1 over the best Multi-layer.

For the IEMOCAP-Environment (Park), the best Single-layer representation (Decoder, best-performing layer) achieved 67.74\% in UAR and 67.47\% in F1. The best Multi-layer fusion approach (Encoder, Weighted Sum) reached 70.56\% in UAR and 69.43\% in F1. The proposed Layer Adapter achieved 73.00\% in UAR and 72.42\% in F1, corresponding to improvements of 5.26\% in UAR and 4.95\% in F1 over the best Single-layer, and 2.44\% in UAR and 2.99\% in F1 compared with the best Multi-layer.

For the IEMOCAP-Environment (Traffic), the best Single-layer representation (Encoder, best-performing layer) obtained 69.86\% in UAR and 69.93\% in F1. The strongest Multi-layer fusion method (Encoder, Weighted Sum) achieved 72.21\% in UAR and 71.87\% in F1. The proposed Layer Adapter further improved performance to 74.39\% in UAR and 74.04\% in F1, providing relative gains of 4.53\% in UAR and 4.11\% in F1 compared with the best Single-layer, and 2.18\% in UAR and 2.17\% in F1 over the best Multi-layer.

Taken together, these results demonstrate that while both Single-layer and Multi-layer fusion strategies capture valuable ASR features, the Layer Adapter method, by jointly leveraging encoder and decoder representations, consistently delivers superior performance through integrating complementary information across layers and modules.

\vspace{-2mm}
\subsection{Layer-wise Analysis of ASR and SSL Representations for NSER}	

\begin{figure*}[htbp]
    \centering   
    \includegraphics[width=\linewidth]{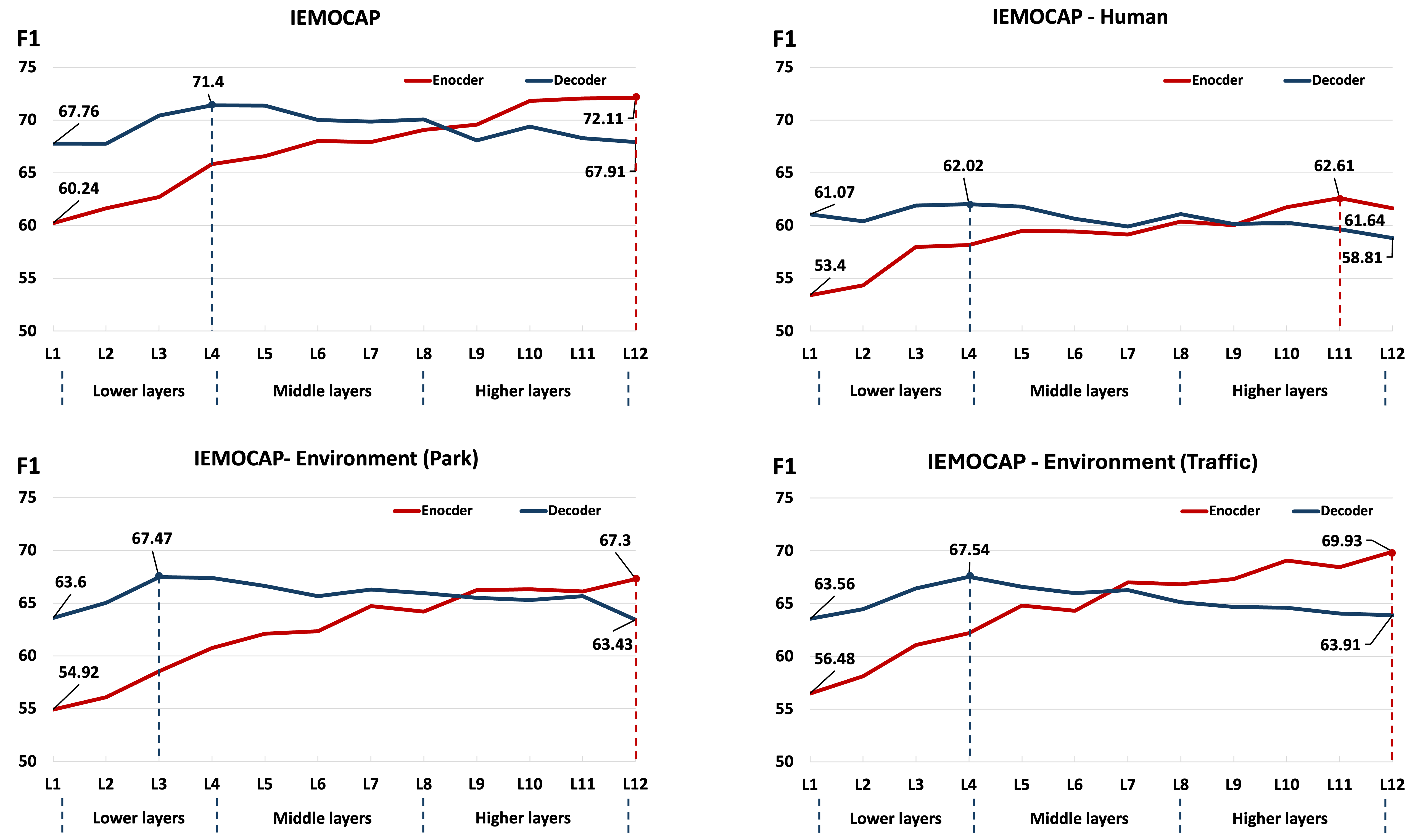}
    \caption{Layer-wise performance (F1) of ASR (Whisper) representations.}
\end{figure*}

\begin{figure}[t]
     \centering
     \includegraphics[width=\linewidth]{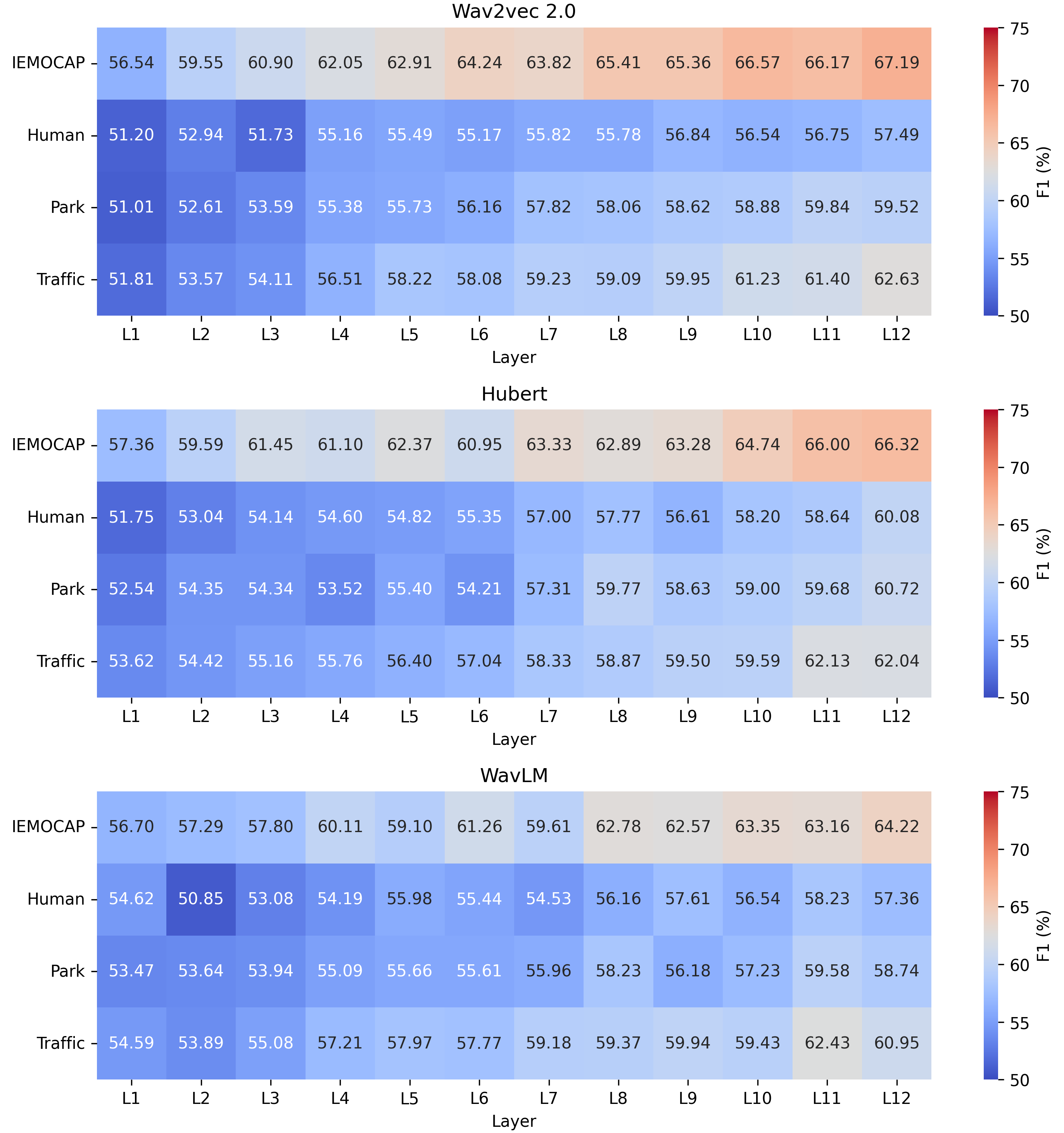}
     \caption{Layer-wise (F1) heatmaps of SSL models.}
\vspace{-5mm}
\end{figure}

To further investigate how different representation levels contribute to NSER, we group the layers into three categories: low (L1–L4), middle (L5–L8), and high (L9–L12). As illustrated in Fig. 3 and Fig. 4, the performance of the ASR encoder steadily improves with increasing depth, reaching its peak in the high-layer group. In contrast, the ASR decoder achieves its best performance in the middle layers before declining. For all SSL encoders (Wav2Vec 2.0, HuBERT, and WavLM), performance improves progressively across layers and generally peaks in the high-layer group, similar to the ASR encoder.

For the IEMOCAP dataset, the ASR encoder improves steadily, starting from 60.24\% at L1 and reaching a peak of 72.11\% at L12 in the high-layer group. In contrast, the ASR decoder achieves its best performance in the middle layer group (71.40\% at L4) before declining to 67.91\% at L12. SSL models rise more gradually and reach lower peaks: Wav2Vec 2.0 attains 67.19\% at L12, HuBERT 66.32\% at L12, and WavLM 64.22\% at L12. Even the strongest SSL model lags behind the Whisper encoder and decoder by at least 4\%, highlighting that ASR representations capture richer semantic cues in clean speech.

In the Human subset, the encoder performance increases steadily from 53.40\% at L1 to a peak of 62.61\% at L11 in the high-layer group, whereas the decoder peaks earlier at 62.20\% (L4) before declining to 58.81\% at L12. Among SSL models, HuBERT performs best with 60.08\% at L12, followed by WavLM (59.32\%, L11), while Wav2Vec 2.0 lags behind at 57.49\% (L12). Although the performance gap between the best SSL model and the Whisper encoder is narrower in this subset, SSL models still underperform overall, indicating that ASR representations remain more robust to human noise.

In the Environment–Park subset, the encoder performance increases from 54.92\% at L1 to a peak of 67.30\% at L12 in the high-layer group, whereas the decoder peaks earlier at 67.47\% (L3) before declining to 63.43\% at L12. Among SSL models, HuBERT performs best with 60.72\% at L12, followed by Wav2Vec 2.0 (59.84\%, L11) and WavLM (59.58\%, L11). Although these results approach the decoder’s performance, they remain more than 6\% below the encoder, further underscoring the superior robustness of ASR-based representations under fluctuating park noise.

In the more challenging Environment–Traffic subset, the encoder performance increases from 56.48\% at L1 to a peak of 69.93\% at L12 in the high-layer group, whereas the decoder peaks earlier at 67.54\% (L4) and then steadily declines to 63.91\% at L12. Among SSL models, Wav2Vec 2.0 performs best with 62.63\% at L12, followed closely by WavLM (62.43\%, L11) and HuBERT (62.13\%, L11). Although these results approach the decoder’s performance, they remain more than 6\% below the encoder, underscoring the superior robustness of ASR-based representations under fluctuating traffic noise.

Overall, the layer-wise analysis highlights a clear division of strengths: higher Whisper encoder layers provide dominant semantic and noise-robust features, while middle decoder layers supply complementary phonetic–semantic alignment. SSL models such as Wav2Vec 2.0, HuBERT, and WavLM exhibit gradual improvements and peak at higher layers, with HuBERT and WavLM generally outperforming Wav2Vec 2.0 in noisy subsets, yet all SSLs remain consistently below the encoder and decoder.

\subsection{Robustness of ASR Representations on NSER}	
To evaluate the robustness of ASR representations for NSER, we compared the encoder (Whisper-Enc), the decoder (Whisper-Dec), and their integration (Proposed) against state-of-the-art SSL models across different SNR levels. The results are presented in Table V.

\begin{table*}[htbp]
\centering
\caption{Exploring the Robustness of ASR Representations on NSER for IEMOCAP across different SNR levels.}
\renewcommand{\arraystretch}{1.5}
\setlength{\tabcolsep}{8pt}
%\scriptsize
\centering
\resizebox{\linewidth}{!}{
\begin{tabular}{c|c|cc|cc|cc|cc|cc}
\toprule
\multirow{2}{*}{\textbf{Model}}
& \multirow{2}{*}{\textbf{Feature}}
& \multicolumn{2}{c|}{\textbf{SNR -10 dB}}
& \multicolumn{2}{c|}{\textbf{SNR -5 dB}}
& \multicolumn{2}{c|}{\textbf{SNR 0 dB}}
& \multicolumn{2}{c|}{\textbf{SNR 5 dB}}
& \multicolumn{2}{c}{\textbf{SNR 10 dB}} \\
\cline{3-12}
& & \textbf{UAR(\%)} & \textbf{F1(\%)} & \textbf{UAR(\%)} & \textbf{F1(\%)} & \textbf{UAR(\%)} & \textbf{F1(\%)} & \textbf{UAR(\%)} & \textbf{F1(\%)} & \textbf{UAR(\%)} & \textbf{F1(\%)} \\
\midrule
\midrule
\multicolumn{12}{c}{\textbf{IEMOCAP - Human}} \\
\midrule
\midrule
\multirow{3}{*}{SSL} 
& Wav2Vec 2.0 (FT) & 45.63 & 45.22 & 50.26 & 49.26 & 59.24 & 57.49 & 64.40 & 62.66 & 65.83 & 64.80\\
& HuBERT (FT) & 51.43 & 50.22 & 54.62 & 54.57 & 61.08 & 60.08 & 64.40 & 62.90 & 65.57 & 64.48 \\
& WavLM (FT) & 53.43 & 52.03 & 56.63 & 54.33 & 59.28 & 57.36 & 61.65 & 60.89 & 64.23 & 63.92 \\
\hline
\multirow{3}{*}{ASR}
& Whisper-Enc & 61.21 & 60.54 & 61.45 & 60.77 & 67.04 & 66.36 & 70.08 & 70.22 & 71.27 & 70.87 \\
& Whisper-Dec & 60.20 & 59.09 & 60.42 & 60.19 & 66.99 & 66.61 & 68.38 & 68.78 & 69.97 & 69.94 \\
\cline{2-12}
& \textbf{Proposed} & \textbf{61.92} & \textbf{60.87} & \textbf{63.06} & \textbf{61.58} & \textbf{68.79} & \textbf{68.39} & \textbf{70.81} & \textbf{70.71} & \textbf{73.35} & \textbf{73.02} \\
\midrule
\midrule
\multicolumn{12}{c}{\textbf{IEMOCAP - Environment (Park)}} \\
\midrule
\midrule
\multirow{3}{*}{SSL} 
& Wav2Vec 2.0 (FT) & 52.73 & 52.03 & 57.35 & 56.41 & 60.68 & 59.52 & 65.02 & 64.19 & 65.15 & 64.75 \\
& HuBERT (FT) & 53.10 & 51.21 & 57.71 & 56.21 & 61.41 & 60.72 & 64.53 & 63.94 & 65.56 & 64.70 \\
& WavLM (FT) & 53.76 & 52.65 & 56.17 & 54.40 & 60.97 & 58.74 & 63.24 & 62.49 & 63.83 & 62.63 \\
\hline
\multirow{3}{*}{ASR} 
& Whisper-Enc & 64.55 & 63.29 & 68.36 & 67.58 & 70.56 & 69.43 & 73.14 & 72.66 & 73.64 & 73.16 \\
& Whisper-Dec & 61.36 & 59.82 & 67.20 & 65.99 & 69.72 & 68.39 & 70.07 & 68.99 & 70.42 & 69.31 \\
\cline{2-12}
& \textbf{Proposed} & \textbf{65.37} & \textbf{64.39} & \textbf{68.78} & \textbf{67.94} & \textbf{73.00} & \textbf{72.42} & \textbf{73.84} & \textbf{73.41} & \textbf{74.83} & \textbf{74.38} \\
\midrule
\midrule
\multicolumn{12}{c}{\textbf{IEMOCAP - Environment (Traffic)}} \\
\midrule
\midrule
\multirow{3}{*}{SSL} 
& Wav2Vec 2.0 (FT) 
& 55.08 & 53.06 & 60.88 & 59.17 & 64.12 & 62.63 & 65.89 & 64.88 & 66.56 & 65.59 \\
& HuBERT (FT) 
& 55.29 & 53.93 & 59.71 & 58.59 & 62.66 & 62.04 & 65.15 & 64.09 & 67.12 & 65.86 \\
& WavLM (FT) 
& 55.54 & 54.23 & 59.35 & 57.69 & 62.28 & 60.95 & 65.26 & 64.15 & 65.55 & 64.41 \\
\hline
\multirow{3}{*}{ASR}
& Whisper-Enc & 67.65 & 66.82 & 71.42 & 71.29 & 72.21 & 71.87 & 72.94 & 72.81 & 73.72 & 73.25 \\
& Whisper-Dec & 66.34 & 65.97 & 68.12 & 68.18 & 71.90 & 69.12 & 69.54 & 69.42 & 70.23 & 69.41 \\
\cline{2-12}
& \textbf{Proposed} & \textbf{69.11} & \textbf{67.92} & \textbf{72.79} & \textbf{72.45} & \textbf{74.39} & \textbf{74.04} & \textbf{74.86} & \textbf{74.61} & \textbf{75.30} & \textbf{74.93} \\
\bottomrule
\end{tabular}}
\end{table*}

\begin{figure*}[t]
     \centering
     \includegraphics[width=\linewidth]{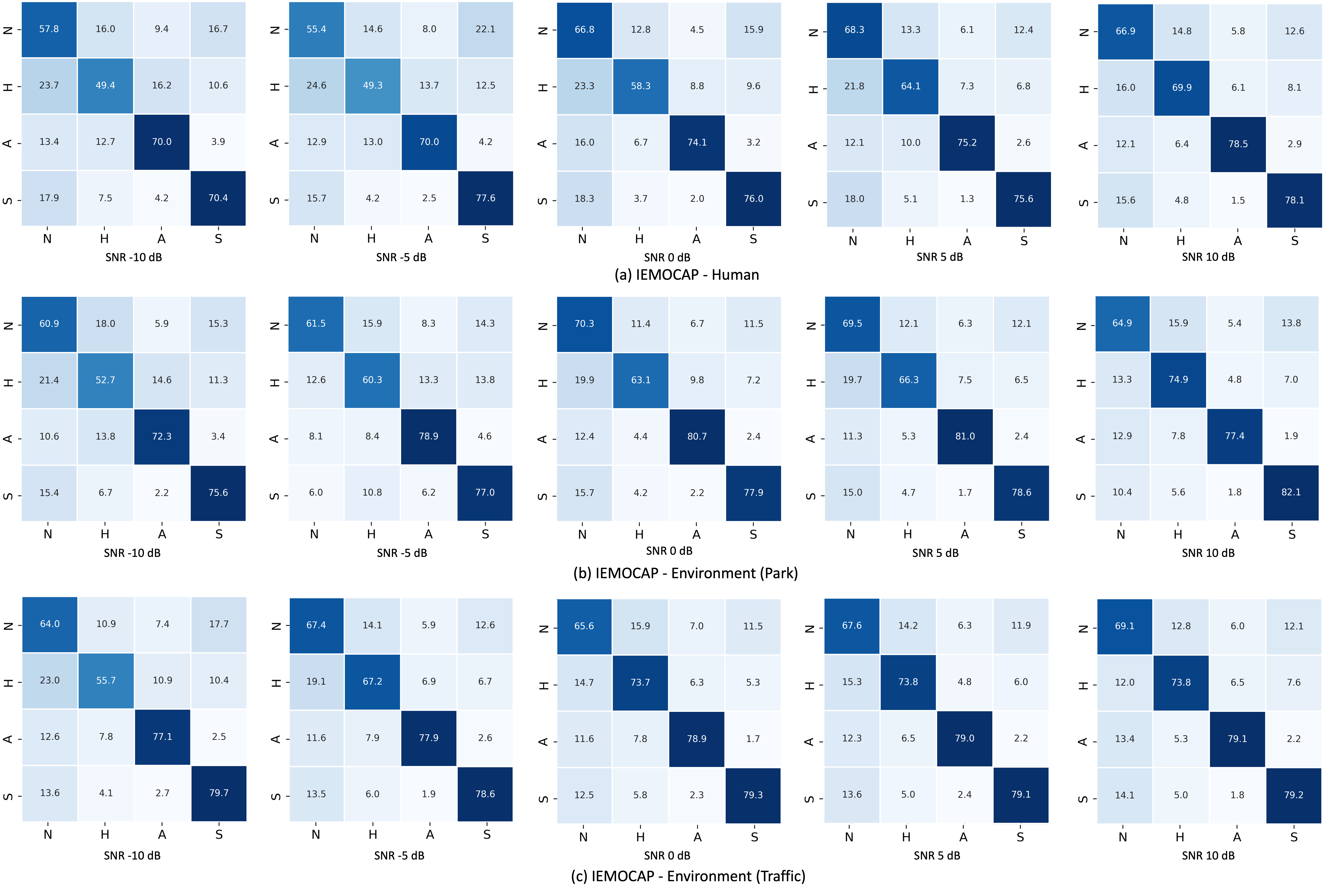}
     \caption{Class-wise robustness analysis under varying SNR conditions. Abbreviations of class labels are as follows: N (Neutral), H (Happy), A (Angry), and S (Sad).}
\vspace{-5mm}
 \end{figure*}
 
In the Human subset, performance degradation is most pronounced under extreme noise. At –10 dB, the proposed method achieves 61.92\% UAR and 60.87\% F1. As SNR increases to 0 dB, performance improves by 6.87\% in UAR and 7.52\% in F1. At 10 dB, gains to –10 dB reach 11.43\% in UAR and 12.15\% in F1, with the method achieving 73.35\% UAR and 73.02\% F1. For the SSL baseline (Wav2Vec 2.0), UAR and F1 rise from 45.63\% and 45.22\% at –10 dB to 65.83\% and 64.80\% at 10 dB, corresponding to improvements of 20.20\% and 19.58\%. Nevertheless, at 10 dB it still trails the proposed method by 7.52\% in UAR and 8.22\% in F1, underscoring the robustness of encoder–decoder integration under both extremely noisy and relatively clean conditions.

In the IEMOCAP-Environment (Park) subset, at –10 dB, the proposed method achieves 65.37\% UAR and 64.39\% F1. As the SNR increases to 0 dB, performance improves by 7.63\% in UAR and 8.03\% in F1. At 10 dB, the gains relative to –10 dB reach 9.46\% in UAR and 9.99\% in F1, with the method achieving 74.83\% UAR and 74.38\% F1. For the SSL baseline (Wav2Vec 2.0), UAR and F1 increase from 52.73\% and 52.03\% at –10 dB to 65.15\% and 64.75\% at 10 dB, corresponding to improvements of 12.42\% in UAR and 12.72\% in F1. However, at 10 dB it still trails the proposed method by 9.68\% in UAR and 9.63\% in F1, confirming the superiority of the proposed approach across all noise levels.

In the IEMOCAP-Environment (Traffic) subset, at –10 dB, the proposed method achieves 69.11\% UAR and 67.92\% F1. As the SNR increases to 0 dB, performance improves by 5.28\% in UAR and 6.12\% in F1. At 10 dB, the gains relative to –10 dB reach 6.19\% in UAR and 7.01\% in F1, with the method achieving 75.30\% UAR and 74.93\% F1. For the SSL baseline (HuBERT), UAR and F1 rise from 55.29\% and 53.93\% at –10 dB to 67.12\% and 65.86\% at 10 dB, corresponding to improvements of 11.83\% in UAR and 11.93\% in F1. Nevertheless, at 10 dB it still falls short of the proposed method by 8.18\% in UAR and 9.07\% in F1, highlighting the resilience of ASR-based representations under challenging traffic noise conditions.

In addition, Fig. 5 provides a heatmap visualization of the confusion matrices across different SNR conditions, offering deeper insights into category-level robustness. A consistent observation is that Angry remains the most resilient class, with accuracy above 70\% even under extreme –10 dB noise, suggesting that its prosodic and acoustic cues are less affected by background interference. In contrast, Happy exhibits the highest sensitivity to noise, often being misclassified as Neutral, particularly at lower SNR levels (–10 dB and –5 dB). This indicates that subtle emotional expressions are more vulnerable to degradation in noisy environments. Meanwhile, Sad shows relatively stable performance.

Overall, across both Human and Environmental subsets, the proposed model consistently demonstrates superior robustness compared with SSL-based features, even under extreme –10 dB noise conditions. These results confirm the proposed method as a more reliable and noise-resilient representation for NSER across diverse realistic acoustic environments.

\subsection{Correlation between ASR Performance and NSER Performance}	

Fig. 6 illustrates the correlation between ASR performance and NSER performance, examining the effects of different noise intensities and types on the effectiveness of ASR. The analysis incorporates representations extracted from Whisper, representing a large-scale ASR approach, and from a Conformer-based ASR model, representing a conventional ASR approach, to compare their ASR and NSER performances under various noisy conditions.

For the IEMOCAP dataset Whisper attains a WER of 0.28 and an NSER of 76.53$\%$  for F1. Conversely, Conformer records a WER of 0.55 and an NSER of 73.04$\%$  for F1, showing slightly lower robustness compared to Whisper.

In the IEMOCAP-Human subset, performance degradation is more pronounced. For the Whisper model, at 5 dB SNR, the WER is 0.36. This WER increases to 0.57 at SNR of 0 dB and further rises to 0.86 at -5 dB, indicating a substantial decline in performance with decreasing SNR. Concurrently, the F1 score for NSER decreases by 9.13$\%$ as the noise level increases from 5 to -5 dB. For the Conformer model, performance worsens consistently across all noise levels, with the WER increasing to 0.65 at 5 dB, rising further to 0.75 at 0 dB, and peaking at 0.86 at -5 dB. Similarly, NSER sees a 6.01$\%$ decrease in F1 as noise levels increase from 5 to -5 dB.

In the IEMOCAP-Environment (Park) subset, the ASR performance gradually deteriorates as the SNR decreases. For Whisper, at an SNR of 5 dB, the WER increases to 0.32. At an SNR of 0 dB, this metric rises to 0.39. Under the most challenging condition of -5 dB, the WER peaks at 0.54. Additionally, the F1 score for NSER tasks decreases by 5.47$\%$ as the noise level increases from 5 to -5 dB, indicating degraded performance in NSER tasks under higher noise conditions. For Conformer, performance shows greater sensitivity to noise, with WER increasing to 0.75 at 5 dB SNR, rising further to 0.86 at 0 dB SNR, and reaching 0.95 at -5 dB SNR. The F1 score for NSER using the Conformer model decreases by 8.4$\%$ as the noise level increases from 5 to -5 dB, indicating that the Conformer model's performance deteriorates under higher-noise-level conditions.

\begin{table*}[htbp]
\centering
\caption{Impact of ASR representations in speech and text modalities on NSER.}
\renewcommand{\arraystretch}{1.5}
\setlength{\tabcolsep}{8pt}
\resizebox{\linewidth}{!}{
\begin{tabular}{c|c|cc|cc|cc|cc|cc} 
\toprule
\multirow{3}{*}{\textbf{Modality}} & \multirow{3}{*}{\textbf{Model}} 
& \multicolumn{2}{c|}{\multirow{2}{*}{\textbf{MELD}}}
& \multicolumn{2}{c|}{\multirow{2}{*}{\textbf{IEMOCAP}}}
& \multicolumn{2}{c|}{\multirow{2}{*}{\makecell{\textbf{IEMOCAP}\\\textbf{(Human)}}}}
& \multicolumn{4}{c}{\makecell{\textbf{IEMOCAP}\\\textbf{(Environment)}}} \\ 
\cline{9-12}
& & \multicolumn{2}{c|}{} & \multicolumn{2}{c|}{} & \multicolumn{2}{c|}{}
  & \multicolumn{2}{c|}{\textbf{Park}} & \multicolumn{2}{c}{\textbf{Traffic}} \\
\cline{3-12}
& & \textbf{UAR(\%)} & \textbf{F1(\%)} 
  & \textbf{UAR(\%)} & \textbf{F1(\%)} 
  & \textbf{UAR(\%)} & \textbf{F1(\%)} 
  & \textbf{UAR(\%)} & \textbf{F1(\%)} 
  & \textbf{UAR(\%)} & \textbf{F1(\%)} \\ 
\midrule
\midrule
\multirow{2}{*}{Text} 
& BERT (Whisper)   & 40.79 & 42.31 & 67.43 & 67.26 & 56.87 & 56.51 & 62.91 & 63.03 & 64.65 & 64.86 \\
& BERT (Transcripts)  & 43.18 & 45.50 & 70.09 & 69.59 & \textbf{70.09} & \textbf{69.59} & 70.09 & 69.59 & 70.09 & 69.59 \\
\hline
\multirow{3}{*}{Speech}  
& Whisper-Enc  & 46.71 & 48.01 & 75.02 & 74.10 & 67.04 & 66.36 & 70.56 & 69.43 & 72.21 & 71.87 \\
& Whisper-Dec  & 46.37 & 47.55 & 72.75 & 72.39 & 66.99 & 66.61 & 69.72 & 68.39 & 71.90 & 69.21 \\
\cline{2-12}
& \textbf{Proposed}  
& \textbf{48.95} & \textbf{48.87} 
& \textbf{76.50} & \textbf{76.53} 
& 68.79 & 68.39
& \textbf{73.00} & \textbf{72.42} & \textbf{74.39} & \textbf{74.04}\\
\bottomrule
\end{tabular}}
\end{table*}

\begin{figure}[htbp]
    \centering
    \includegraphics[width=\linewidth]{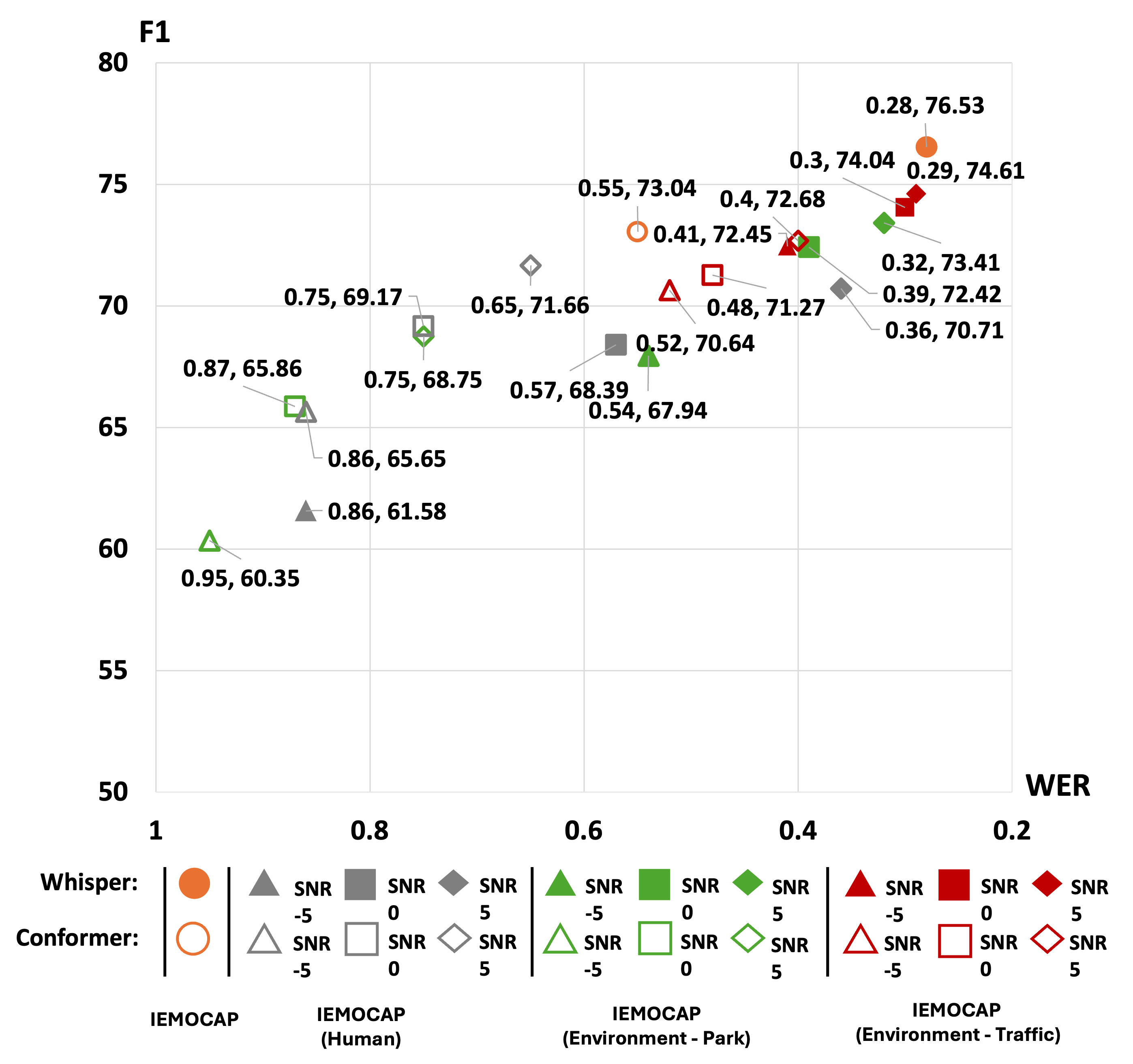}
    \caption{Correlation between ASR (WER) and NSER (F1) performances using Whisper and Conformer (Pearson’s correlation coefficient $r=-0.89$, $p<0.001$).}
\vspace{-5mm}
\end{figure}

In the IEMOCAP-Environment (Traffic) subset, ASR performance degrades consistently with decreasing SNR. For the Whisper model, the WER increases from 0.29 at 5 dB to 0.30 at 0 dB and further to 0.41 at -5 dB. Correspondingly, the NSER F1 score decreases by 2.16$\%$  as the SNR drops from 5 to -5 dB, indicating reduced robustness under severe traffic noise. For the Conformer model, degradation is more pronounced, with the WER rising from 0.40 at 5 dB to 0.48 at 0 dB and 0.52 at -5 dB. Similarly, the NSER F1 score drops by 2.04$\%$, revealing a higher vulnerability to traffic noise compared with Whisper.

In summary, these results emphasize the significant impact of noise on ASR performance, particularly in scenarios with complex human noise. A consistent correlation between ASR performance and NSER is evident within the same datasets, i.e., across the IEMOCAP-Human subset and the IEMOCAP-Environment subsets (Park and Traffic), when comparing different ASR systems. Furthermore, NSER strongly depends on the noise type, with human noise exerting a more disruptive effect than environmental noise. Moreover, as noise levels increase, both ASR models exhibit deteriorating performance on NSER tasks. This decrease is slightly less pronounced in Whisper, indicating its comparative robustness in handling diverse noisy conditions, including both human and environmental noise. Overall, aggregating across all datasets and both ASR models, a strong negative correlation between ASR performance (WER) and NSER performance (F1) is observed (Pearson’s $r=-0.89$, $p<0.001$).

\vspace{-2mm}
\subsection{Impact of ASR Representations in Speech and Text Modalities on NSER}	

To investigate the impact of ASR representations in speech and text modalities on NSER, we conduct a comparative analysis, contrasting different text inputs for BERT in the
text modality with various ASR representations in the speech modality, as detailed in Table VI.

In the text modality, BERT applied to transcripts consistently outperforms BERT on Whisper-generated text, with the largest margin observed on the IEMOCAP-Human subset. However, across most datasets, the speech modality demonstrates superior performance compared with the text modality, except for the IEMOCAP-Human subset, highlighting the added value of ASR representations for NSER.

For the MELD dataset, the Proposed method achieves 48.95\% UAR and 48.87\% F1, improving over BERT (Transcripts) by 5.77\% in UAR and 3.37\% in F1. The Encoder alone improves by 3.53\% in UAR and 2.51\% in F1, while the Decoder adds 3.19\% in UAR and 2.05\% in F1.

For the IEMOCAP dataset, the Proposed method reaches 76.50\% UAR and 76.53\% F1, yielding improvements of 6.41\% in UAR and 6.94\% in F1 over BERT (Transcripts). The Encoder contributes 4.93\% in UAR and 4.51\% in F1, while the Decoder provides 2.66\% in UAR and 2.80\% in F1.

For the IEMOCAP-Human subset, BERT (Transcripts) achieves 70.09\% UAR and 69.59\% F1, which remains higher than the speech-based alternatives. The Proposed method obtains 68.79\% UAR and 68.39\% F1, falling short by 1.30\% in UAR and 1.20\% in F1 compared with BERT (Transcripts). The Encoder lags by 3.05\% in UAR and 3.23\% in F1, while the Decoder is lower by 3.10\% in UAR and 2.98\% in F1.

For the IEMOCAP-Environment (Park) subset, the Proposed method reaches 73.00\% UAR and 72.42\% F1, surpassing BERT (Transcripts) by 2.91\% in UAR and 2.83\% in F1. The Encoder provides smaller gains of 0.47\% in UAR and a slight decrease of 0.16\% in F1, while the Decoder remains lower by 0.37\% in UAR and 1.20\% in F1.

For the IEMOCAP-Environment (Traffic) subset, the Proposed method achieves 74.39\% UAR and 74.04\% F1, outperforming BERT (Transcripts) by 4.3\% in UAR and 4.45\% in F1. The Encoder independently improves by 2.12\% in UAR and 2.28\% in F1, while the Decoder contributes 1.81\% in UAR but falls behind by 0.38\% in F1.

Overall, these findings show that ASR-based speech representations generally outperform text-based representations for NSER, particularly under environmental and mixed noise conditions where robustness is critical. However, in the Human subset, BERT on clean transcripts remains slightly stronger, suggesting that text modality can still provide advantages when high-quality transcriptions are available. This highlights the complementary roles of speech and text modalities in NSER.

\subsection{Robustness of ASR Representations in Scenarios Involving Cross-lingual NSER}	
To evaluate the robustness of ASR representations in cross-lingual NSER tasks, we conducted experiments where neither the ASR nor SSL models were exposed to the target language during training. A monolingual ASR model was used as a baseline. Whisper served as the ASR model, providing both encoder and decoder representations. Table VII summarizes the performance of SSL models, Mel-Spectrogram, and Whisper-based ASR representations.

\begin{table}[htbp]
\tiny
\centering
\caption{Impact of the cross-lingual setting on NSER performance using Mel-Spectrograms, SSL representations (\textbf{FT}: fine-tuning, \textbf{FZ}: frozen), and ASR representations (Whisper-Enc/Dec in \textbf{EN}: English, \textbf{CN}: Chinese).}
\renewcommand{\arraystretch}{1.5} % 调整行距
\resizebox{\linewidth}{!}{
\begin{tabular}{c|c|cc} 
\toprule
\textbf{Source} & \textbf{Model} & \textbf{UAR(\%)} & \textbf{F1(\%)}  \\ 
\midrule
\midrule
\begin{tabular}[c]{@{}c@{}}Mel- \\Spectrogram\end{tabular} & BiGRU & 53.53 & 52.91 \\ 
\hline
\multirow{12}{*}{Waveform} 
                        & Wav2Vec 2.0 (FZ) & 22.00 & 18.79 \\
                        & HuBERT (FZ) & 26.49 & 22.88 \\                   
                        & WavLM (FZ) & 21.47 & 18.80 \\
\cline{2-4}
                        & Wav2Vec 2.0 (FT) & 52.44 & 52.39 \\
                        & HuBERT (FT) & 51.44 & 50.88 \\
                        & WavLM (FT) & 53.89 & 52.17 \\
\cline{2-4}
                        & Whisper-Enc(EN) & 52.90 & 52.84 \\
                        & Whisper-Dec(EN) & 55.30 & 54.26 \\
                        & Whisper-Enc+Dec(EN) & 55.90 & 54.52 \\
\cline{2-4}
                        & Whisper-Enc(CN) & 56.43 & 55.25 \\
                        & Whisper-Dec(CN) & 57.41 & 56.71 \\
                        & Whisper-Enc+Dec(CN) & \textbf{58.54} & \textbf{58.13} \\
\bottomrule
\end{tabular}}
\end{table}

The results reveal substantial differences in the performance of various representations. Mel-Spectrogram results serve as a benchmark for comparison. In contrast, SSL models under a frozen configuration show marked performance degradation. For instance, frozen HuBERT's performance is lowered by 27.04$\%$ in UAR and 30.03$\%$ in F1 compared with the Mel-Spectrogram. These findings highlight the inability of frozen SSL representations to generalize effectively in cross-lingual scenarios. However, through fine-tuning, the performance of SSL models improves significantly. For example, WavLM shows increases of 32.42$\%$ in UAR and 33.37$\%$ in F1 compared with its frozen configuration.

Whisper-based ASR representations exhibit notably stronger robustness, particularly in cross-lingual (EN) settings. Compared with the Mel-Spectrogram baseline, Whisper-Decoder yields additional gains of 1.77$\%$ in UAR and 1.35$\%$ in F1. The Whisper-Encoder+Decoder achieves its largest gains with a 2.37$\%$ increase in UAR and a 1.61$\%$ improvement in F1 over the baseline, illustrating the effectiveness of leveraging complementary features from both modules for cross-lingual emotion recognition.

To further clarify the impact of language mismatch, we compared Whisper-EN and Whisper-CN representations under cross-lingual and monolingual scenarios. Relative to Whisper-CN, Whisper-EN shows declines in both UAR and F1 across all components. For the Encoder, there is a decrease of 3.53$\%$ in UAR and 2.41$\%$ in F1, whereas the Decoder experiences reductions of 2.11$\%$ in UAR and 2.45$\%$ in F1. The Encoder+Decoder shows the largest decline, with 2.64$\%$ lower UAR and 3.61$\%$ lower F1, highlighting the compounded challenges of language mismatch in cross-lingual settings.

Overall, Whisper-based ASR representations consistently surpass both SSL and Mel-Spectrogram baselines in cross-lingual and monolingual contexts, demonstrating remarkable robustness against language mismatch and noise. Moreover, given that CASIA provides parallel utterances with identical lexical content across different emotions, the observed improvements can be attributed primarily to paralinguistic cues rather than lexical factors. In particular, the consistent gains of Whisper-based ASR representations over Mel-Spectrogram and SSL models suggest that their advantage stems not only from semantic encoding, but also from a superior ability to capture emotion-relevant prosody, intonation, and spectral dynamics across diverse conditions.

\section{Conclusions and Future Work}
In this paper, we proposed a novel approach to NSER by leveraging ASR models as robust feature extractors in noisy environments. Our study demonstrated its effectiveness in addressing the limitations of conventional NSER techniques under noise challenges in realistic acoustic environments. By utilizing intermediate-layer representations from ASR models, we captured emotional speech features with greater accuracy, resulting in superior NSER performance compared with conventional noise reduction methods and SSL approaches.

In our thorough analysis, we examined the effects of ASR representations across several factors, including noise intensity, types of noise, and modality differences. The experimental results revealed key insights: the proposed method consistently outperforms conventional and SSL approaches, showing robustness against various noise intensities and types, and even surpasses text-based approaches using ASR transcriptions. Notably, our study highlighted the relevance of cross-lingual scenarios, where our method sustains robust performance compared with self-supervised representations.

Looking ahead, future work will further build upon the empirical findings of this study by exploring more effective ASR-based NSER frameworks from multiple perspectives. First, given the observed performance variations across different Whisper layers, adaptive or task-aware layer selection strategies could be investigated to more explicitly exploit the complementary roles of encoder and decoder representations. Second, the negative correlation between ASR performance and NSER performance suggests that ASR quality may serve as a useful reference for optimizing downstream emotion recognition under noisy conditions, motivating the design of more tightly coupled ASR–SER co-optimization frameworks. Finally, evaluations will be extended to broader datasets and robustness across different languages and cultural contexts will be explored to further validate the generalization and practical applicability of the proposed framework.

\bibliographystyle{IEEEtran}
\bibliography{refs}

%\newpage
 

%\bf{If you include a photo:}\vspace{-33pt}
%\begin{IEEEbiography}[{\includegraphics[width=1in,height=1.25in,clip,keepaspectratio]{fig1}}]{Michael Shell}
%Use $\backslash${\tt{begin\{IEEEbiography\}}} and then for the 1st argument use $\backslash${\tt{includegraphics}} to declare and link the author photo.Use the author name as the 3rd argument followed by the biography %text.
%\end{IEEEbiography}



\vfill

\end{document}